\documentclass[aps,pre,reprint]{revtex4-1}
\usepackage{setspace}
\usepackage{amsfonts,amsmath,amssymb}
\usepackage{txfonts}
\usepackage{graphicx}
\usepackage{bm}
\usepackage{epstopdf}
\usepackage{verbatim}
\usepackage{units}
\usepackage[T1]{fontenc}
\usepackage[latin9]{inputenc}
\usepackage{amsbsy}
\usepackage{esint}

\begin{document}

\title{Electromotive force due to magnetohydrodynamic fluctuations in sheared rotating
turbulence}

\author{J.~Squire}
\email{jsquire@princeton.edu}
%\affiliation{Department of Astrophysical Sciences and Princeton Plasma Physics Laboratory, Princeton University,Princeton, NJ 08543}
\author{A.~Bhattacharjee}
\affiliation{Max Planck/Princeton Center for Plasma Physics, Department of Astrophysical Sciences and Princeton Plasma Physics Laboratory, Princeton University, Princeton, NJ 08543, USA}

\begin{abstract}
This article presents a calculation of the mean electromotive force arising from 
general small-scale magnetohydrodynamical turbulence, within the framework of the second-order
correlation approximation. With the goal of improving understanding of the accretion
disk dynamo, effects arising through small-scale magnetic fluctuations, velocity gradients, density 
and turbulence stratification, and rotation, are included. The primary result, which
supplements numerical findings, is that an off-diagonal turbulent resistivity due 
to \emph{magnetic fluctuations} can produce large-scale dynamo action -- the magnetic analogue 
of the ``shear-current'' effect. In addition, consideration of $\alpha$ effects in the stratified regions of
disks gives the puzzling result that there is no strong prediction for a sign of $\alpha$, since the effects 
due to kinetic and magnetic fluctuations, as well as those due to shear and rotation, are each of opposing signs and tend to cancel each other.
\end{abstract}
\maketitle

\section{Introduction}

% ADD URPIN CITATIONS
% ADD PARKER AND TOBIAS CITATIONS
Explaining the amplification of magnetic fields with correlation lengths larger than the 
underlying  fluid motions has proven to be a fascinating and rich
problem in astrophysics. From the early days of mean-field dynamo 
theory it has been well known that the presence of fluid helicity 
enables such behavior \cite{Moffatt:1978tc,Krause:1980vr}. This is the so-called $\alpha$ effect, where 
the small-scale turbulence creates an electromotive force (EMF) 
$$\bm{\mathcal{E}}=\left< \bm{u}\times \bm{b} \right>$$
that is proportional to a large-scale magnetic field, $\bm{\mathcal{E}}=\alpha \bm{B}$,
leading to exponential instability in the kinematic regime. While this simple $\alpha$ effect is now well established 
and regularly observed in simulations, a variety of complications
exist in explaining observations. For one, in some situations -- for instance,
the inner regions of accretion disks -- there is no reason to expect a helical flow and 
symmetry arguments demonstrate that $\alpha=0$, 
yet dynamo action is still observed in numerical experiments \cite{Hawley:1995gd,Lesur:2008cv}. 
Less obviously, nonlinear effects caused by the fast build up of 
small-scale fields can ``quench'' $\alpha$ dynamos
before significant mean-field amplitudes are reached \cite{Kulsrud:1992ej,Gruzinov:1994ea}. 
Since the effectiveness of this quenching increases with the Reynolds numbers,
it remains unclear whether mean-field theory is able to explain the observed
field amplitudes in the nearly dissipation-free plasmas prevalent in astrophysical environments.
For these reasons, it is interesting to consider other possibilities
for mean-field dynamo action, in particular the effects of velocity gradients and
strong homogenous magnetic fluctuations.

In this paper, we present a very general theoretical
examination of different mean-field dynamo effects, within the
second-order correlation approximation (SOCA). In particular,
we include the effects of specified large-scale velocity gradients, rotation, 
density and turbulence stratification, helicity, and a bath
of strong small-scale magnetic fluctuations (treated in the same
way as the velocity fluctuations). For our primary inspiration
in this work -- the accretion disk dynamo -- each of these effects can be 
important in some way, and this will also be the case in a wide variety 
of other astrophysical scenarios. Of particular note is the presence of homogenous magnetic 
fluctuations, which have not been included in most previous theoretical 
mean-field dynamo investigations 
(but see, for example, Refs.~\cite{Vainshtein:1983dq,Rogachevskii:2004cx,Radler:2003gg,Pipin:2008hx}).
These should
be generically present, at a similar level to velocity fluctuations, in magnetohydrodynamic (MHD) turbulence 
above moderate Reynolds numbers, due to small-scale
dynamo action. While SOCA itself cannot capture the 
small-scale dynamo, by assuming the presence of
the magnetic fluctuations we can compute expected changes to the EMF, in particular 
whether a small-scale magnetic field might suppress, or enhance, kinematic dynamo effects.

The most important result presented here is an analytic 
confirmation of our numerical work related to the ``magnetic shear-current effect'' \cite{LowRm,HighRm}. Generically, 
this type of dynamo is non-helical, driven by 
the interaction of an off-diagonal turbulent resistivity with 
a mean shear flow \cite{Moffatt:1982ta,Urpin:2002ct,Urpin:1999wl,Rogachevskii:2003gg}.  
Some controversy has surrounded the kinematic version
of this effect, since following early work \cite{Urpin:1999wl,Urpin:2002ct,Rogachevskii:2003gg}, 
others found that the crucial transport coefficient $\eta_{yx}$ had the incorrect
sign to promote dynamo action \cite{Rudiger:2006gx,Radler:2006hy,Brandenburg:2008bc,SINGH:2011ho}.
Here, we show that the magnetic version of this effect is much more robust and of the correct sign --  
not only is its magnitude substantially larger than the kinematic effect,
but a variety of calculation methods agree on this: SOCA, the 
spectral $\tau$~approximation \cite{Rogachevskii:2004cx},  quasi-linear theory \cite{SINGH:2011ho,LowRm},
and perturbative shearing wave calculations \cite{Lesur:2008fn}.
With this array of other calculations, we feel that SOCA calculations
are important, not because they should be more accurate than other methods, but
because they are simple, have a well-understood range of validity, and allow exploration of 
expressions across a range of parameters (e.g., magnetic Prandtl number). This final consideration is 
notable since it provides the 
researcher with some indication of the robustness of a given effect, 
for instance by noting if the sign a given transport coefficient is particularly sensitive 
to slight changes in parameters. 
Finally, all of our results related to $\eta_{yx}$ have been 
confirmed through direct numerical simulations \cite{LowRm,HighRm}.
Most important is the measurement of a marked decrease in $\eta_{yx}$ 
after saturation of the small-scale dynamo in sheared turbulence, accompanied by excitation
of a coherent mean-field dynamo \cite{HighRm}.

Turbulence and density stratification is invariably  
significant in astrophysical scenarios, including in 
accretion disks away from the central plane of the disk. 
With this application in mind, we also apply our results 
to the case of stratified rotating turbulence with strong velocity shear, 
considering the resulting $\alpha$ effects. We
find that for a Keplerian (or more generally, anticyclonic) rotation profile, the contributions
from shear and rotation, and those from kinetic and magnetic fluctuations, 
are each of opposite signs. The dominant contribution  
will depend strongly on the magnetic Prandtl number $\mathrm{Pm}$, as well as 
the relative intensities of magnetic and kinetic turbulence. 
This is confusing in light of the beautifully
coherent ``butterfly diagrams'' that are often seen in stratified accretion-disk
 simulations \cite{Brandenburg:2002cia,Brandenburg:1995dc,Gressel:2010dj,Simon:2012dq},
which would suggest a robust negative value for $\alpha_{yy}$. 
We note that the contributions to these $\alpha$ effects from 
velocity shear are at least as strong as those from rotation and should 
not generally be neglected. 

The structure of our calculation almost identically follows that 
of \citet{Radler:2006hy}  (hereafter\defcitealias{Radler:2006hy}{RS06} RS06), with the additional effects of magnetic fluctuations, density stratification 
(within an anelastic approximation) and net helicity. The inclusion
of such a variety of physical effects leads to a rather prodigious
number of terms, and we have used the \emph{VEST} package \cite{Squire:2014vq} 
in \emph{Mathematica} to carry out the bulk of the calculations. 
We start, in Sec.~\ref{sec:fundamentals}, by outlining the setup of the calculation, 
including the most general form of $\bm{\mathcal{E}}$ allowed by the symmetries of the problem,
as well as the relation of the transport coefficients in Cartesian domains 
with velocity shear to this general form. We also give the perturbation 
expansion used, which
is a generalization of that in \citetalias{Radler:2006hy} to include magnetic turbulence
at lowest order. 
In Sec.~\ref{sec:Outline}, we outline the procedure used in the calculation 
itself, skipping many details for the sake of brevity.  Particular focus is placed on the 
unstratified shear dynamo -- especially the magnetic
shear-current effect -- in Sec.~\ref{sec:Specific-results-for}, while 
the stratified $\alpha$ effect is examined in the same geometry in Sec.~\ref{sec:stratified}. Readers
interested primarily in the application of calculated coefficients to disk
dynamos may wish to skip directly to these sections. 
Due to the length of algebraic expressions, the full set
of transport coefficients is given in Appendix~\ref{sec:List-of-all transports coeffs}.

\section{Fundamentals of mean-field electrodynamics}\label{sec:fundamentals}

Our starting point, common to most mean-field dynamo calculations,
is the system of compressible MHD equations,
\begin{align}
\frac{\partial\rho}{\partial t}  &+ \nabla \cdot (\rho \bm{U}_{T})=0, \nonumber \\[1.5ex]
\rho\frac{\partial\bm{U}_{T}}{\partial t}  &+\rho\left(\bm{U}_{T}\cdot\nabla\right)\bm{U}_{T}+2\rho\bm{\Omega}\times\bm{U}_{T}+\nabla p= \bm{B}_{T}\cdot\nabla\bm{B}_{T}\nonumber \\
 &\quad+\nabla\cdot [ \rho \nu (\nabla \bm{U}_{T}+(\nabla \bm{U}_{T})^{T})+\rho \bar{\zeta} \delta_{ij}\nabla \cdot \bm{U}_{T} ]
 +\bm{\sigma}_{\bm{u}},\nonumber \\[1.5ex]
\frac{\partial\bm{B}_{T}}{\partial t} & =\nabla\times\left(\bm{U}_{T}\times\bm{B}_{T}\right)+\eta\nabla^{2}\bm{B}_{t}+\bm{\sigma}_{\bm{b}},\nonumber \\[1.5ex]
 & \nabla\cdot\bm{U}_{T}=0,\;\;\;\nabla\cdot\bm{B}_{t}=0.\label{eq:MHD}
\end{align}
Here $\bm{U}_{T}$ and $\bm{B}_{T}$ are the full velocity and magnetic
fields, $\bar{\nu}$~is the kinematic viscosity, $\bar{\zeta}$~is the bulk viscosity (this will not contribute), and $\bar{\eta}$~is the resistivity. 
We have included the effects of rotation through a mean Coriolis force ($2\rho \bm{\Omega}\times\bm{U}_{T}$)
in the momentum equation. Before calculating transport coefficients
from Eq.~\eqref{eq:MHD}, we shall apply an anelastic approximation 
\cite{Kichatinov:1992vg,Rudiger:1993vx}, assuming nearly
incompressible fluctuations with $\nabla \cdot (\rho \bm{u})=0$ [see Eq.~\eqref{eq:MFdecomposition}]. This
allows  low-order effects due to a mean density gradient to be retained, while still 
preserving most of the simplicity of an incompressible calculation. 

Mean-field dynamo theory \citep{Moffatt:1978tc,Krause:1980vr}
involves splitting fields into a mean and fluctuating part; 
\begin{equation}
\bm{U}_{T}=\bm{U}+\bm{u},\,\,\, \bm{B}_{T}=\bm{B}+\bm{b},\label{eq:MFdecomposition}
\end{equation}
with $\bm{U}=\left\langle \bm{U}_{T}\right\rangle$, 
$\bm{B}=\left\langle \bm{B}_{T}\right\rangle $. The averaging operation
$\left\langle \cdot\right\rangle $ should filter out small scales
and satisfy the Reynolds averaging rules (later in the manuscript
we will specify $\left\langle \cdot\right\rangle $ as a horizontal
spatial average). Applying $\left\langle \cdot\right\rangle $ to
the induction equation leads to the well-known mean-field induction
equation
\begin{equation}
\partial_{t}\bm{B}=\nabla\times\left(\bm{U}\times\bm{B}\right)+\nabla\times\bm{\mathcal{E}}+\nu\triangle\bm{B}.\label{eq:genMF}
\end{equation}
where $\bm{\mathcal{E}}=\left\langle \bm{u}\times\bm{b}\right\rangle $
is the electromotive force (EMF). The goal of mean-field theory is
to calculate $\bm{\mathcal{E}}$ as a function of $\bm{B}$ and other parameters
in the problem (i.e., $\bm{U}$, $\Omega$, $\nabla \ln \rho$ and the small-scale turbulence
statistics), thereby closing Eq.~\eqref{eq:genMF}. If $\bm{\mathcal{E}}\left(\bm{B}\right)$
is such that a small magnetic field will be reinforced by the small-scale
turbulence,  a dynamo instability results. 

Before commencing with a full calculation of $\bm{\mathcal{E}}$, it is
worth examining the symmetries of the problem. Assuming scale separation
between the mean and fluctuating fields, we can Taylor expand the
EMF as 
\begin{equation}
\mathcal{E}_{i}=a_{ij}B_{j}+b_{ijk}B_{j,k}+\dots
\end{equation}
where we use the Einstein summation convention and the comma denotes
a derivative. The tensors $a_{ij}$ and $b_{ijk}$ are the transport
coefficients determined by the turbulence. In keeping with the separation
of scales assumption, we shall consider linear $\bm{B}$ fields $\left(\bm{B}\right)_{i}=B_{i}+B_{ij}x_{j}$,
velocity fields $\left(\bm{U}\right)_{i}=U_{ij}x_{j}$ and density $\rho = \rho_{0}+\rho_{0}\,\bm{x}\!\cdot\!\nabla\ln\rho$ 
(the constant velocity part can be removed by Galilean  transformation).
As in \citetalias{Radler:2006hy}, 
to cleanly separate different dynamo
effects into scalar coefficients, it is helpful to split $\nabla\bm{U}$ and $\nabla\bm{B}$
into symmetric and antisymmetric parts,
\begin{gather}
U_{ij}=D_{ij}-A_{ij}^{U}=D_{ij}-\frac{1}{2}\varepsilon_{ijk}W_{k},\nonumber \\
B_{ij}=\left(\nabla\bm{B}\right)_{ij}^{(s)}-A_{ij}^{B}=\left(\nabla\bm{B}\right)_{ij}^{(s)}-\frac{1}{2}\varepsilon_{ijk}J_{k},\label{eq:sym spiltting}
\end{gather}
where $D_{ij}$ and $\left(\nabla\bm{B}\right)_{ij}^{(s)}$ are the
symmetric and antisymmetric parts of $U_{ij}$ and $B_{ij}$, $\bm{W}=\nabla\times\bm{U}$
is the background vorticity and $\bm{J}=\nabla\times\bm{B}$ is the
mean current. Due to the assumption $\nabla \cdot \bm{U}=0$ in our calculation,
we have implicitly assumed $\bm{U}\cdot \nabla\rho=0$, a requirement
that could  easily be relaxed if desired.   

We consider general inhomogenous background
turbulence in both $\bm{u}$ and $\bm{b}$, modified by 
mean velocity gradients, rotation and density stratification. The density 
stratification is assumed to be aligned with the turbulence
stratification in the direction $\hat{\bm{g}}$, but we allow 
their magnitudes and signs to differ; that is, defining
\begin{equation}
\nabla \ln \rho =\chi_{\rho} \hat{\bm{g}},\:
\nabla \ln \bar{u} =\chi_{ \bar{u}} \hat{\bm{g}},\:\nabla \ln \bar{b} =\chi_{ \bar{b}} \hat{\bm{g}},
\end{equation}
(where $u_{\mathrm{rms}}=\langle u_{0}^{2}\rangle^{1/2}$, $b_{\mathrm{rms}}=\langle b_{0}^{2}\rangle^{1/2}$), we allow 
$\chi_{ \rho}\neq \chi_{ \bar{u}} \neq \chi_{ \bar{b}}$.
For completeness, 
we include both non-helical and helical contributions to the turbulence \footnote{Our primary reasoning for including the helical part of the correlation
here has been to check that  standard results, e.g., $\alpha^{(0)}\sim\left< \bm{u}\cdot\nabla\times\bm{u}\right> -\left< \bm{b}\cdot\nabla\times\bm{b}\right> $ \cite{Pouquet:1976eg} are obtained using this method.}
but neglect the effects of inhomogeneity on the helical part
\footnote{Stratification of  helical 
turbulence would presumably provide a host of contributions to the 
resistivity tensor that would likely be much smaller than contributions
from the nonhelical fluctuations, due to being higher order. Given the rather immense size of the 
calculation without such effects, it seemed prudent to ignore these.}.
We assume that the EMF due to the background turbulence
vanishes, $\left\langle \bm{u}\times\bm{b}\right\rangle _{0}=0$.
Such a $\bm{B}$ independent contribution could be important in some
situations (see, for example, \citet{Yoshizawa:1993dv}) and
the method applied here can be used to calculate well-known effects of this type if desired, for 
instance the cross-helicity effect \citep{Yokoi:2013di}. In addition, 
we do not calculate the components of the Reynolds stress, which would
force a mean-field velocity $\bm{U}$. This is not justified for any particular
reason other than our primary interest in the magnetic field dynamics. While it is possible that  
there are important interactions between $\bm{U}$ and $\bm{B}$ that lead to other instabilities
\cite{Courvoisier:2010bj}, we leave their systematic study to future work. 

A careful consideration of the symmetry properties of the system leads
to the general representation of $\bm{\mathcal{E}}$ in terms of a set
of scalar transport transport coefficients (see \citetalias{Radler:2006hy}
for a full explanation)
\begin{align}
\bm{\bm{\mathcal{E}}} =&-\alpha^{(0)}_{H}\bm{B}-\alpha^{(D)}_{H} {D}_{ij} {B}_{j} -\gamma^{(\Omega )}_{H} \bm{\Omega}\times \bm{B} - \gamma^{(W )}_{H} \bm{W}\times \bm{B}\nonumber \\-
&\alpha_{1}^{(\Omega)}(\hat{\bm{g}}\cdot \bm{\Omega})\bm{B}-\alpha_{2}^{(\Omega)} [(\hat{\bm{g}}\cdot \bm{B})\bm{\Omega}+(\bm{B}\cdot \bm{\Omega})\hat{\bm{g}}] \nonumber \\-
&\alpha_{1}^{(W)}(\hat{\bm{g}}\cdot \bm{W})\bm{B}-\alpha_{2}^{(W)} [(\hat{\bm{g}}\cdot \bm{B})\bm{W}+(\bm{B}\cdot \bm{W})\hat{\bm{g}}]\nonumber \\-
&\alpha^{(D)}(\varepsilon_{ilm}D_{lj}\hat{g}_{m}+\varepsilon_{jlm}D_{li}\hat{g}_{m})B_{j} \nonumber \\ -
&(\gamma^{(0)}+\gamma^{(\Omega)}\hat{\bm{g}}\times\bm{\Omega}+\gamma^{(W)}\hat{\bm{g}}\times\bm{W}+\gamma^{(D)}D_{ij}\hat{g}_{j})\times\bm{B} \nonumber \\ -
 & \beta^{(0)}\bm{J}-\beta^{(D)}{D}_{ij}{J}_{j}-\left(\delta^{(W)}\bm{W}+\delta^{(\Omega)}\bm{\Omega}\right)\times\bm{J}\nonumber \\ -
 & \left(\kappa^{(W)}\bm{W}+\kappa^{(\Omega)}\bm{\Omega}\right)_{j}\left(\nabla\bm{B}\right)^{(s)}_{ji}-2\kappa^{(D)}\varepsilon_{ijk}D_{kr}\left(\nabla\bm{B}\right)_{jr}^{(s)}\label{eq:general E}
\end{align}
Here we have conformed to the sign conventions in \citetalias{Radler:2006hy} and use 
the Einstein summation convention. The
subscript $\cdot_{H}$ denotes a coefficient
that is only allowed by the helical part of the turbulence, while
all other coefficients arise only 
through the nonhelical part. In addition, since we assume small-scale 
fluctuations in both $\bm{u}$
and $\bm{b}$, we further split each transport coefficient into these
contributions; e.g., $\kappa^{(W)}=(\kappa^{(W)})_{u}+(\kappa^{(W)})_{b}$
Since we work with SOCA in the linear regime (where $\bm{B}$ is small), 
these are always additive and can be calculated separately from the $\bm{u}$ and $\bm{b}$
turbulent contributions. 

\subsection{Cartesian domains}

In Sec.~\ref{sec:Specific-results-for} we shall give specific results
for the numerically convenient Cartesian shear dynamo with nonhelical, unstratified
background turbulence. This is essentially a generalization of
the unstratified shearing box that is often used in accretion-disk simulations. In this case, mean fields depend only on $z$,
$\bm{U}=-Sx\hat{\bm{y}}$ (giving $\bm{W}=-S\hat{\bm{z}}$), $\bm{\Omega}=\Omega\hat{\bm{z}}$
and the mean-field average is defined as an average over $x$ and
$y$, $\left\langle \cdot\right\rangle =(L_{x}L_{y})^{-1}\int\cdot\, dxdy$.
The mean-field equations simplify to
\begin{gather}
\partial_{t}B_{x}=-\eta_{yx}\partial_{z}^{2}B_{y}+\eta_{yy}\partial_{z}^{2}B_{x},\nonumber \\
\partial_{t}B_{y}=-SB_{x}-\eta_{xy}\partial_{z}^{2}B_{x}+\eta_{xx}\partial_{z}^{2}B_{y},\label{eq:SC sa eqs}
\end{gather}
where the $\eta_{ij}$ are defined to be the relevant components of
$b_{ijk}$ that are nonzero for the chosen average and mean field.
For $B_{i}=B_{i0}e^{ikz}e^{\Gamma t}$ a coherent dynamo is possible if 
\begin{equation}
\Gamma=k\left\{-S\eta_{yx}+k^{2}\left[\eta_{xy}\eta_{yx}+\frac{1}{2}(\eta_{xx}-\eta_{yy})^{2}\right]\right\}^{1/2}-k^{2}(\eta_{xx}+\eta_{yy})\label{eq:SC growth}
\end{equation}
has a real part greater than 0. One can neglect the term multiplying $k^{2}$
in the square root in Eq.~\eqref{eq:SC growth} since $S$ is presumed
to be large compared to all transport coefficients. This gives $\eta_{21}S<0$
as a necessary condition for instability. Computing the relationship
between Eq.~\eqref{eq:general E} and Eq.~\eqref{eq:SC sa eqs}
shows
\begin{gather}
\eta_{yx}=-S\left[\delta^{(W)}-\frac{1}{2}\left(\kappa^{(W)}-\beta^{(D)}+\kappa^{(D)}\right)\right]+\Omega\left(\delta^{(\Omega)}-\frac{1}{2}\kappa^{(\Omega)}\right),\nonumber \\
\eta_{xy}=S\left[\delta^{(W)}-\frac{1}{2}\left(\kappa^{(W)}+\beta^{(D)}-\kappa^{(D)}\right)\right]-\Omega\left(\delta^{(\Omega)}-\frac{1}{2}\kappa^{(\Omega)}\right),\label{eq: eta from beta}
\end{gather}
and $\eta_{xx}=\eta_{yy}=\beta^{(0)}$. Note that Eq.~\eqref{eq:SC growth}
only describes the growth due to a coherent dynamo process and fluctuations
in $\alpha$ or $\eta$ that arise in any finite system can cause
a dynamo in and of themselves \citep{Heinemann:2011gt,Mitra:2012iv,LowRm}.
We shall specialize to the Cartesian case in Sec.~\ref{sec:Specific-results-for}
and keep $\bm{U}$ general for the calculation of the transport coefficients listed in Eq.~\eqref{eq:general E}.

In Sec.~\ref{sec:stratified} we give results specific to the case of stratified 
sheared rotating turbulence. This is motivated by consideration of 
the upper (or lower) portions of an accretion disk. Again, mean fields 
depend only on $z$, $\bm{U}=-Sx\hat{\bm{y}}$, $\bm{\Omega}=\Omega\hat{\bm{z}}$,
and $\hat{\bm{g}}=\hat{\bm{z}}$. We neglect off-diagonal
resistivity contributions and use $\eta_{xx}=\eta_{yy}=\beta^{(0)}$.
The mean-field equations simplify to
\begin{gather}
\partial_{t}B_{x}= -a_{yx}\partial_{z}B_{x}-a_{yy}\partial_{z}B_{y}+\beta^{(0)}\partial_{z}^{2}B_{x}\nonumber \\
\partial_{t}B_{y}=-SB_{x}+a_{xy}\partial_{z}B_{x}+a_{xx}\partial_{z}B_{y}+\beta^{(0)}\partial_{z}^{2}B_{y}.\label{eq:alpha eqs}
\end{gather}
With $a_{xy}=-a_{yx}$, considering $B_{i}=B_{i0}e^{ikz}e^{\Gamma t}$, one obtains the growth rate
\begin{align}
\Gamma =&\left(i k S  a_{yy} /2+k^{2} a_{yy} a_{xx}\right)^{1/2}+ika_{xy}-k^2 \beta^{(0)}  
.\label{eq:growth alpha}
\end{align}
Again, $S$ is presumed large in comparison to all transport coefficients, 
so we see that any nonzero $a_{yy}$ can lead to instability at sufficiently long wavelength. 
Of course, in practice there will be a minimum $k$ possible in the system, particularly since 
$a_{yy }$ arises from a stratification, so a finite $a_{yy}$ will be necessary 
to overcome the turbulent resistivity. The coefficients in Eq.~\eqref{eq:alpha eqs} are 
related to those in Eq.~\eqref{eq:general E} through $a_{xy}=-a_{yx}=\gamma^{(0)}$ and
\begin{gather}
a_{yy}=S\left(\alpha^{(W)}_{1}-\alpha^{(D) }\right) - \Omega\, \alpha^{(\Omega)}_{1} ,\nonumber\\
a_{xx}=S\left(\alpha^{(W)}_{1}+\alpha^{(D) }\right) - \Omega\, \alpha^{(\Omega)}_{1}.\label{eq:ayy def}
\end{gather}

\subsection{Perturbation expansion to describe the fluctuations}

For the calculation of $\bm{\mathcal{E}}$ we use the second-order correlation 
approximation (SOCA), which involves solving linear equations for the fluctuations
by neglecting third-order and higher correlations. As such, this is
rigorously valid only at low Reynolds numbers 
 where dissipation dominates
over nonlinearities for the fluctuations (SOCA 
can also be valid in the small Strouhal number limit [Eq.~\eqref{eq:DIMENSIONLESS variables}], see \citet{Brandenburg:2005kla}
for a more thorough discussion). In addition, 
we choose to include the shear, rotation and density stratification perturbatively \cite{Rogachevskii:2003gg,Radler:2006hy}, considering
only the linear response of transport coefficients to these effects. 
An analytic
calculation with shear included at zeroth order can be found in \cite{SINGH:2011ho},
and some examples of calculations that include 
nonlinear contributions from other effects can be found in Refs.~\citep{Rudiger:1990de,Kichatinov:1992vg,Rudiger:1993vx,Radler:2003gg,Rogachevskii:2004cx}. In a very general calculation,
 \citet{Pipin:2008hx} nonlinearly  includes all effects discussed here  (although the approach, the ``minimal $\tau$~approximation,'' has 
 a somewhat unknown range of validity).
We have also computed the magnetic dynamo transport coefficients with 
non-perturbative shear and rotation using statistical simulation in the shearing box
\citep{LowRm}. 

Following \citet{Rudiger:1993vx}, \citet{Kichatinov:1992vg}, and \citet{Rudiger:1990de}, we 
start by making an anelastic approximation to the full compressible equations,
$\nabla \cdot (\rho \bm{u}) =0$. This should be valid for weakly compressible turbulence
and allows the inclusion of a weak density stratification into the problem, which 
is important in a wide variety of mean-field dynamos. We shall
assume that the large-scale flow is incompressible, since 
our primary application is to shear flows. It is then
more convenient to work in terms of the small-scale momentum
\cite{Rudiger:1993vx,Kichatinov:1992vg}, $\bm{m}\equiv \rho \bm{u}$, 
since the calculation for $\bm{m}$ proceeds in a similar manner to 
the incompressible case. 

In retaining both strong homogenous velocity and magnetic fluctuations, denoted
$\bm{u}_{0}$ (or $\bm{m}_{0}$) and $\bm{b}_{0}$ respectively, we must treat the momentum
and induction equations on the same theoretical footing. We start from Eq.~\eqref{eq:MHD} 
by splitting into mean-field and fluctuation equations, 
applying the anelastic approximation followed by the change of variables $\bm{u}_{0}=\bm{m}_{0}/\rho$. We then linearize 
the small-scale equations
and expand $\bm{m}=\bm{m}_{0}+\bm{m}^{(0)}+\bm{m}^{(1)}\dots$, 
$\bm{b}=\bm{b}_{0}+\bm{b}^{(0)}+\bm{b}^{(1)}\dots$, to perturbatively 
find the change to the background turbulence caused by the 
shear, rotation and stratification. This leads the SOCA equations that will 
be used to calculate all transport coefficients:
\begin{align}
\left(\partial_{t}-\nu\triangle\right)\bm{m}^{(0)}  &=-\left(\bm{m}_{0}\cdot\nabla\bm{U}+\bm{U}\cdot\nabla\bm{m}_{0} - (\bm{g}_{\rho}\cdot \bm{U}) \bm{m}_{0} \right)\nonumber \\ -\nabla p^{(0)}-2&\bm{\Omega}\times\bm{m}_{0}+
 \left(\bm{b}_{0}\cdot\nabla\bm{B}+\bm{B}\cdot\nabla\bm{b}^{(0)}\right) -\nu\bm{g}_{\rho}\cdot\nabla \bm{m}_{0} \nonumber \\
\left(\partial_{t}-\nu\triangle\right)\bm{m}^{(1)}  &=-\left(\bm{m}^{(0)}\cdot\nabla\bm{U}+\bm{U}\cdot\nabla\bm{m}^{(0)} - (\bm{g}_{\rho}\cdot \bm{U}) \bm{m}^{(0)} \right)\nonumber \\ -\nabla p^{(1)}-2&\bm{\Omega}\times\bm{m}^{(0)}+
 \left(\bm{b}^{(0)}\cdot\nabla\bm{B}+\bm{B}\cdot\nabla\bm{b}^{(0)}\right) -\nu\bm{g}_{\rho}\cdot\nabla \bm{m}^{(0)} \nonumber \\
 \left(\partial_{t}-\eta\triangle\right)\bm{b}^{(0)} & = \rho^{-1} \left[(\bm{g}_{\rho}\cdot \bm{m}_{0}) \bm{B} -\bm{m}_{0}\cdot \nabla \bm{B} +\bm{B}\cdot \nabla \bm{m}_{0} \right. \nonumber \\
-&\left.(\bm{g}_{\rho}\cdot \bm{B})   \bm{m}_{0}\right] +\bm{b}_{0}\cdot \nabla \bm{U} -\bm{U}\cdot \nabla \bm{b}_{0}, \nonumber \\
 \left(\partial_{t}-\eta\triangle\right)\bm{b}^{(1)} & = \rho^{-1} \left[(\bm{g}_{\rho}\cdot \bm{m}^{(0)}) \bm{B} -\bm{m}^{(0)}\cdot \nabla \bm{B} +\bm{B}\cdot \nabla \bm{m}^{(0)} \right. \nonumber \\
-& \left.(\bm{g}_{\rho}\cdot \bm{B})  \bm{m}^{(0)}\right] +\bm{b}^{(0)}\cdot \nabla \bm{U} -\bm{U}\cdot \nabla \bm{b}^{(0)}, \label{eq:perturbation expansion}
\end{align}
along with divergence constraints for each $\bm{m}^{(0)},\,\bm{b}^{(0)},\,\bm{m}^{(1)}$, and $\bm{b}^{(1)}$.
Here $\bm{g}_{\rho}\equiv \chi_{\rho}\hat{\bm{g}} $ and we have neglected second 
derivatives of $\bm{U}$ and $\rho$, as well as products of $\nabla \bm{B}$ with $\chi_{\rho}$ [these
contributions should vanish in the transport coefficients, since the Eq.~\eqref{eq:general E} 
illustrates that there is no contribution to the resistivity due to $\hat{\bm{g}}$ at linear order]. 
In addition, we shall neglect any terms that involve quadratic products of $\bm{U}$, $\bm{\Omega}$, 
and $\chi_{\rho}$ (e.g., $(\bm{g}_{\rho}\cdot \bm{U}) \bm{m}_{0}$), and expand all terms to linear 
order to take the Fourier transport of Eq.~\eqref{eq:perturbation expansion} (see App.~\ref{app:Fourier space}).

While it may seem surprising that one requires terms two orders higher
than $\bm{m}_{0}$ and $\bm{b}_{0}$, it is straightforward to see
that only considering $\bm{m}^{(0)}$ and $\bm{b}^{(0)}$ will not
lead to contributions to $\bm{\mathcal{E}}$ that depend on products of
$\bm{B}$ with $\bm{U}$ or $\bm{\Omega}$ (these are the interesting
terms in the dynamo, describing the effect of rotation or velocity).
With this in mind, the EMF is calculated as
\begin{align}
\mathcal{E}_{ij}= & \left\langle u_{i}b_{j}\right\rangle =\left\langle \rho^{-1} m_{0i}b_{0j}\right\rangle +\left\langle \rho^{-1} m_{0i}b_{j}^{(0)}\right\rangle +\left\langle \rho^{-1} m_{0i}b_{j}^{(1)}\right\rangle \nonumber \\+
 & \left\langle \rho^{-1}m^{(0)}_{i}b_{0j}\right\rangle +\left\langle \rho^{-1}m_{i}^{(1)} b_{0j}\right\rangle +\left\langle \rho^{-1}m_{i}^{(0)}b_{j}^{(0)}\right\rangle .\label{eq:EMF expansion}
\end{align}
Despite the fact that all the terms in Eq.~\eqref{eq:EMF expansion}
give some contribution, there are also a large number of terms that
contain quadratic products of $U_{ij}$, $\Omega_{i}$, $\chi_{\rho}$, or $\bm{B}$, which are
neglected. As is
evident, with background turbulence in both $\bm{u}$ and $\bm{b} $ there will be contributions
to $\bm{\mathcal{E}}$ from the Maxwell stress ($\bm{B}\cdot\nabla\bm{b}+\bm{b}\cdot\nabla\bm{B}$)
that one would expect to be of a similar magnitude to the standard
kinematic dynamo arising from the Lorentz force {[}$\nabla\times\left(\bm{u}\times\bm{B}\right)${]}.
This choice of perturbation expansion is the natural generalization
of \citetalias{Radler:2006hy} to the case with $\bm{b}_{0}$ fluctuations
(although note that $\bm{u}^{(1)}$ in \citetalias{Radler:2006hy} has
become $\bm{u}^{(0)}$ in our notation such that $\bm{u}$ and $\bm{b}$
are treated on equal footings). Our results for the kinematic dynamo
($\bm{b}_{0}=0$) without density stratification agree with \citetalias{Radler:2006hy} aside from a single
numerical coefficient (see App.~\ref{sec:List-of-all transports coeffs}).

\section{Outline of the calculation of $\bm{\mathcal{E}}$}\label{sec:Outline}

Our calculation  follows the methods and notation in \citetalias{Radler:2006hy}
and a full explanation is given there. Here we give a very brief outline,
in particular the choices involved, with final results
given in Appendix~\ref{sec:List-of-all transports coeffs}. We have
carried out the entire calculation in \emph{Mathematica} using the
\emph{VEST} package \citep{Squire:2014vq} to handle abstract tensor manipulations
using the Einstein summation convention. 

The two-point correlation of two fields $\bm{v}$ and $\bm{w}$ is
defined as 
\begin{equation}
\phi_{ij}^{(vw)}\left(\bm{x}_{1,}t_{1};\bm{x}_{2},t_{2}\right)=\left\langle v_{i}\left(\bm{x}_{1,}t_{1}\right)w_{j}\left(\bm{x}_{2},t_{2}\right)\right\rangle .
\end{equation}
It is convenient to write such quantities in the variables
\begin{gather}
\bm{R}=\left(\bm{x}_{1}+\bm{x}_{2}\right)/2, \:\:
\bm{r}=\bm{x}_{1}-\bm{x}_{2},\nonumber\\
T=\left(t_{1}+t_{2}\right)/2, \:\:t=t_{1}-t_{2},
\end{gather}
giving \begin{equation}
\phi_{ij}^{(vw)}\left(\bm{R},T;\bm{r},t\right)=\left\langle v_{i} \left(\bm{R}+\frac{\bm{r}}{2},T+\frac{t}{2} \right)\,w_{j}\left(\bm{R}-\frac{\bm{r}}{2},T-\frac{t}{2}\right)\right\rangle .\end{equation}
One then Fourier transforms in the small-scale variable $\bm{r}$
to obtain 
\begin{equation}
\phi_{ij}^{(vw)}\left(\bm{R},T;\bm{r},t\right)=\int d\bm{k}\,d\omega\,\tilde{\phi}_{ij}^{(vw)}\left(\bm{R},T;\bm{k},\omega\right)e^{i\left(\bm{k}\cdot\bm{r}-\omega t\right)},
\end{equation}
with 
\begin{equation}
\tilde{\phi}_{ij}^{(vw)}\left(\bm{R},T;\bm{k},\omega\right)=\int d\bm{K}\,d\Omega\,\left\langle \left[\hat{v}\right]_{+}\left[\hat{w}\right]_{-}\right\rangle e^{i\left(\bm{K}\cdot\bm{R}-\Omega T\right)},
\end{equation}
where $\hat{v}=\hat{v}\left(\bm{k},\omega\right)$ and  $\hat{w}=\hat{w}\left(\bm{k},\omega\right)$
denote the Fourier transforms of $v$ and $w$, and we use the $\left[\cdot\right]_{\pm}$
notation of \citetalias{Radler:2006hy},
\begin{equation}
\left[\hat{f}\left(\bm{k},\omega\right)\right]_{\pm}=\hat{f}\left(\pm\bm{k}+\bm{K}/2,\pm \omega+\Omega/2\right).
\end{equation}
As in \citetalias{Radler:2006hy} we shall calculate 
\begin{align}
&\mathcal{E}_{ij}  \left(\bm{R},T;0,0\right)=\int d\bm{k}\,d\omega\,\tilde{\bm{\bm{\mathcal{E}}}}_{ij}\left(\bm{R},T;\bm{k},\omega\right)\nonumber\\
 & =\int d\bm{K}\,d\Omega\, d\bm{k}\,d\omega\,\left\langle [\rho^{-1}\hat{m}_{i}]_{+}[\hat{b}_{i}]_{-}\right\rangle e^{i\bm{K}\cdot\bm{R}-i\Omega T}  \nonumber \\
&  =\int\,d\bm{K}\,d\Omega\, d\bm{k}\,d\omega\rho_{0}^{-1}\left\langle [\hat{m}_{i} - ig_{\rho j}\partial_{k_{j}}\hat{m}_{i} ]_{+} [\hat{b}_{i}]_{-}\right\rangle e^{i\bm{K}\cdot\bm{R}-i\Omega T},
\end{align}
setting $\bm{R},\, T\rightarrow0$ only after extracting the coefficients
of $B_{i}$ and $B_{ij}$ (i.e., the transport coefficients $a_{ij}$,
$b_{ijk}$). 

With these notations defined, the starting point of the calculation
is the substitution of the linear forms for $\bm{U}$, $\rho$ and $\bm{B}$ and 
into Eq.~\eqref{eq:perturbation expansion}, followed by a Fourier transform.
This leads to Eqs.~\eqref{eq:u0FS}-\eqref{eq:b1FS}. One then substitutes
$\hat{m}_{i}^{(0)}$ and $\hat{b}_{i}^{(0)}$ into $\hat{m}_{i}^{(1)}$ and
$\hat{b}_{i}^{(1)}$ to form explicit expressions for $\hat{u}_{i}$ and
$\hat{b}_{i}$ in terms of $\hat{u}_{0i}$ and $\hat{b}_{i0}$. Defining
\begin{gather*}
\tilde{m}_{ij}=\left\langle [\hat{m}_{0i}]_{+}[\hat{m}_{0j}]_{-}\right\rangle ,\\
\tilde{b}_{ij}=\left\langle [\hat{b}_{0i}]_{+}[\hat{b}_{0j}]_{-}\right\rangle ,
\end{gather*}
to specify the statistics of $\bm{u}_{0}$ and $\bm{b}_{0}$, this
allows one to form Eq.~\eqref{eq:EMF expansion} in terms of $\tilde{m}_{ij}$
and $\tilde{b}_{ij}$, neglecting all terms that contain $U_{ij}U_{rs},$ 
$U_{ij}\Omega_{j}$, $\Omega_{i}\Omega_{r}$, $U_{ij}\chi_{\rho},$ $\Omega_{i}\chi_{\rho},$ $(\nabla \ln\rho )^{2},$ or any products of $B_{i}$
and $B_{ij}$. Recall that we have assumed $\left\langle \bm{u}_{0}\bm{b}_{0}\right\rangle =0$,
implying that all terms in the expansion of $\mathcal{E}_{ij}$ contain
$B_{i}$ or $B_{ij}$. 
In keeping with the expansion to linear order in background quantities, it is necessary 
to expand $[f(\bm{k})]_{\pm}$ to first order in $\bm{K}$ in those terms
that contain $B_{i}$ (i.e., $\alpha$ coefficients). These lead to  terms 
involving the gradient of the turbulence intensity. Note that $[f(\bm{k})]_{\pm}\rightarrow f(\pm\bm{k})$
for resistive terms (coefficients of $B_{ij}$). 
Some useful identities in the above procedure are given
in \citetalias{Radler:2006hy} Eqs.~(33)-(35), which are needed to remove
$\partial/\partial k_{i}$ derivatives from $u_{0i}$ and $b_{0i}$.
Similarly, we apply the identities 
\begin{equation}
k_{i}\tilde{m}_{ij} = -\frac{K_{i}}{2}\tilde{m}_{ij},\quad k_{i}\tilde{m}_{ji} = \frac{K_{i}}{2}\tilde{m}_{ji}
\end{equation}
(and similarly for $\tilde{b}_{ji}$), which arise from the divergence constraints on $\hat{m}_{i}$ and $\hat{b}_{i}$.

Extracting the coefficients of $B_{i}$ and $B_{ij}$ in the expression
for $\mathcal{E}_{i}=\varepsilon_{ijk}\mathcal{E}_{jk}\left(\bm{0},0\right)$,
at this stage we have large integral expressions for $a_{ij}$ and
$b_{ijk}$ in terms of $\tilde{m}_{ij}$ and $\tilde{b}_{ij}$ and their spatial derivatives {[}for
example, \citetalias{Radler:2006hy} Eqs.~(39)-(40){]}. Without further
interpretation, such expressions are nearly useless, and it is helpful
to insert explicit forms for $\tilde{m}_{ij}$ and $\tilde{b}_{ij}$.
Assuming isotropy in the limit of vanishing mean flow
and rotation, we insert 
\begin{align}
\tilde{m}_{ij}=\frac{1}{2}&\left[\delta_{ij}-\frac{k_{i}k_{j}}{k^{2}}-\frac{1}{2k^{2}}(k_{i}K_{j}-k_{j}K_{i})\right]W_{m}\left(\bm{K};{k},\omega\right)\nonumber\\
&\qquad\qquad-i\varepsilon_{ijl}\frac{k_{l}}{k^{2}}H_{u}\left({k},\omega\right),\nonumber \\
\tilde{b}_{ij}=\frac{1}{2}&\left[\delta_{ij}-\frac{k_{i}k_{j}}{k^{2}}-\frac{1}{2k^{2}}(k_{i}K_{j}-k_{j}K_{i})\right]W_{b}\left(\bm{K};{k},\omega\right)\nonumber\\
&\qquad\qquad-i\varepsilon_{ijl}\frac{k_{l}}{k^{2}}H_{b}\left({k},\omega\right),\label{eq:stat forms}
\end{align}
where $k=\left| k_{i} \right|$.
Here $W_{m,b}$ represents
a non-helical part and $H_{m,b}$ a helical part
of the background turbulence \cite{Rudiger:1993vx,Kichatinov:1992vg}. This form 
for $W_{m}$ is particularly convenient since it can be shown that to first order 
in the scale of density variation
\begin{equation}
W_{m}\left(\bm{x};{k},\omega\right) = \rho^{2}(\bm{x}) W_{u}\left(\bm{x};{k},\omega\right),
\end{equation}
where $W_{u}\left(\bm{x};{k},\omega\right)$ is a similar function specifying the
statistics of $\bm{u}$ and $W_{m}\left(\bm{x};{k},\omega\right) =\int d\bm{K} e^{i\bm{K}\cdot \bm{x}}W_{m}\left(\bm{K};{k},\omega\right)$   \cite{Kichatinov:1992vg}. In this way, 
\begin{equation}
\nabla W_{m}\left(\bm{x};{k},\omega\right) = \hat{\bm{g}}(2\chi_{\rho} + 2\chi_{\bar{u}})W_{m}\left(\bm{x};{k},\omega\right),\label{eq:Wm deriv}
\end{equation}
separating the effects due to density and turbulence stratification. 
Similarly, for the magnetic fluctuations
\begin{equation}
\nabla W_{b}\left(\bm{x};{k},\omega\right) = 2\hat{\bm{g}}\chi_{\bar{b}}W_{b}\left(\bm{x};{k},\omega\right).
\end{equation}

It transpires that all terms now depend on $\bm{k}$ only through
$k$, and all of the integrals can be substantially simplified using
\begin{gather}
\int d\bm{k}\,k_{i}k_{j}f\left(k\right)=\frac{1}{3}\delta_{ij}\int dk \,k^{2}f\left(k\right),\nonumber \\
\int d\bm{k} \,k_{i}k_{j}k_{k}k_{l}f\left(k\right)=\frac{1}{15}\left(\delta_{ij}\delta_{kl}+\delta_{ik}\delta_{jl}+\delta_{il}\delta_{jk}\right)\int dk\, k^{4}f\left(k\right),
\label{eq:isotropy integrals}
\end{gather}
where the integrals over $k$ on the right-hand side of Eq.~\eqref{eq:isotropy integrals} are taken from $k=0\rightarrow \infty$.
One then splits $U_{ij}$ and $B_{ij}$ using Eq.~\eqref{eq:sym spiltting},
putting $\mathcal{E}_{i}$ in the form given by Eq.~\eqref{eq:general E}.
One can
straightforwardly read off the transport coefficients $\alpha^{(0)}_{H},\dots,\alpha^{(\Omega)},\dots,\beta^{(0)},\dots$,
as integrals of the form
\begin{gather}
\left(\alpha_{H}^{(\cdot)}\right)_{u,b}=4\pi\int dk\,d\omega\, k^{2}\tilde{\alpha}^{(\cdot)}_{H}\left(k,\omega\right)H_{u,b}\left(k,\omega\right)\nonumber \\
\left(\alpha^{(\cdot)} \right)_{u,b}=4\pi\int dk\,d\omega\, k^{2}\tilde{\alpha}^{(\cdot)}\left(k,\omega\right)W_{u,b}\left(k,\omega\right)\nonumber \\
\left(\beta^{(\cdot)}\right)_{u,b}=4\pi\int dk\,d\omega\,k^{2}\tilde{\beta}^{(\cdot)}\left(k,\omega\right)W_{u,b}\left(k,\omega\right).\label{eq:a and b example}
\end{gather}
 The full list of coefficients $\tilde{\alpha}^{(0)}_{H},\dots,\tilde{\alpha}^{(\Omega)},\dots,\tilde{\beta}^{(0)},\dots$
is given in App.~\ref{sec:List-of-all transports coeffs}.

Finally, it is possible to carry out the integrals of the form in
Eq.~\eqref{eq:a and b example} for a specific form of $W$ and $H$,
leading to explicit expressions for the transport coefficients in
terms of the physical parameters. A convenient form for examining
expressions and plotting is the Gaussian $W$ used in \citetalias{Radler:2006hy},
\begin{equation}
W_{u}=u_{\mathrm{rms}}^{2}\frac{2\lambda_{c}^{3}\tau_{c}}{3\left(2\pi\right)^{5/2}}\frac{\left(k\lambda_{c}\right)^{2}e^{-\left(k\lambda_{c}\right)^{2}/2}}{1+\left(\omega\tau_{c}\right)^{2}},\label{eq:gaussian W}
\end{equation}
with a similar definition of $W_{b}$. With this choice, all integrals
can be carried out explicitly without further approximation. As in
\citetalias{Radler:2006hy}, we shall write such expressions in terms of
the non-dimensional variables (and $\rho_{0}$)
\begin{gather}
\epsilon= b_{\mathrm{rms}}/u_{\mathrm{rms}},\quad p=\lambda_{c}^{2}/\nu\tau_{c},\quad q=\lambda_{c}^{2}/\eta\tau_{c},\quad\mathrm{Pm}=\nu/\eta,\nonumber \\
\mathrm{Re}=u_{\mathrm{rms}}\lambda_{c}/\nu,\quad\mathrm{Rm}=u_{\mathrm{rms}}\lambda_{c}/\eta,\quad\mathrm{St}=u_{\mathrm{rms}}\tau_{c}/\lambda_{c}.\label{eq:DIMENSIONLESS variables}
\end{gather}
Here Pm, Re, Rm, and St are respectively the magnetic Prandtl number,
the fluid Reynolds number, the magnetic Reynolds number and the Strouhal
number. $p$ and $q$ are the ratio of diffusion times, $\lambda_{c}^{2}/\nu$
and $\lambda_{c}^{2}/\eta$, to the correlation time $\tau_{c}$.
Thus $q\rightarrow0$ denotes the low conductivity limit, while $q\rightarrow\infty$
denotes a high conductivity limit (with a similar result for $p$
and fluid diffusivity). A sufficient condition for the validity of
SOCA (i.e., neglect of nonlinear terms in the correlation equations) is
$\mathrm{Rm}\ll1$ in the limit $q\rightarrow0$, and $\mathrm{St}\ll1$
in the limit $q\rightarrow\infty$, see \citet{Brandenburg:2005kla} and \citet{Radler:2006hy}
for more discussion of these validity regimes. In addition, e require $U_{ij}$ and $\Omega_{i}$ be a small perturbation
to the background turbulence.
In practice, we shall use these non-dimensional variables [Eq.~\eqref{eq:DIMENSIONLESS variables}]  for plotting transport coefficients.

We have carried out the full sequence of steps detailed
above in \emph{Mathematica} using the \emph{VEST} package \citep{Squire:2014vq}
to enable straightforward manipulation of tensors in index notation.
This has the obvious advantage of handling the very long expressions
with ease and making the calculation
straightforward to generalize or modify. The sequence of steps is
essentially the same as that detailed above. We first define $\bm{m}^{(0)},$
$\bm{m}^{(1)}$, $\bm{b}^{(0)},$ and $\bm{b}^{(1)}$, insert $\bm{m}^{(0)}$ and
$\bm{b}^{(0)}$ into $\bm{m}^{(1)}$ and $\bm{b}^{(1)}$, then only later
remove products that are quadratic in $U_{ij}$, $\bm{\Omega}$, or $\chi_{\rho}$.
It is then straightforward to define $\left[\cdot\right]_{\pm}$ operators,
their associated product rules, and methods to in expand in $\bm{K}$. This allows the construction
of the entirety of $\bm{\mathcal{E}}$ in one step. Insertion of the explicit
forms for $\tilde{v}_{ij}$ and $\tilde{b}_{ij}$ {[}Eq.~\eqref{eq:stat forms}{]}
and the partial integration using isotropy {[}Eq.~\eqref{eq:isotropy integrals}{]}
is easily carried out using replacement rules. Finally, we decompose
products of $B_{ij}$ with $U_{ij}$, $\bm{\Omega}$ and $\hat{\bm{g}}$ into the form
given in Eq.~\eqref{eq:general E}, allowing the coefficients listed
in App.~\eqref{sec:List-of-all transports coeffs} to be straightforwardly extracted from the total expression. Finally, if so desired, these can be directly
integrated with the specific form of $W$ {[}Eq.~\eqref{eq:gaussian W}{]}
by carefully substituting the dimensionless variables {[}Eq.~\eqref{eq:DIMENSIONLESS variables}{]}
and using \emph{Mathematica}'s native \texttt{Integrate} function.
For the interested reader, we include the full calculation notebook
in supplementary material.

\subsection{Agreement with previous works}

Our results agree with related works of other authors in special limits,
including those utilizing different calculation methods. As discussed
throughout the work, all results of \citetalias{Radler:2006hy} are recovered in
the limit $\nabla \ln \rho=0$
[aside from one discrepancy, in $(\beta^{(D)})_{u}$]. This agrees with \citet{Rudiger:2006gx}, many
results of \citet{Pipin:2008hx}, including his magnetic contributions (see his App.~B),
as well as the quasi-linear methods in \citet{Sridhar:2009jg} and \citet{SINGH:2011ho}.
As is well known, there is a discrepancy between these kinematic quasi-linear
results and those obtained using the $\tau$~approximation \citep{Rogachevskii:2003gg,Rogachevskii:2004cx},
possibly due to a change in sign of $\eta_{yx}$ with $\mathrm{Rm}$ \cite{Brandenburg:2008bc}.
As seen in Eq.~(32) of \citet{Pipin:2008hx}, his conclusions regarding the 
kinetic and magnetic contributions to the shear-current effect (with rotation) are
are similar to ours. 
Our results also compare favorably to previous works without velocity gradients, but
including magnetic fluctuations. As expected, the helical magnetic $\alpha$
effect has the opposite sign to the kinematic effect, and there is
no change to $\beta^{\left(0\right)}$ due to the addition of magnetic
fluctuations. In addition, the signs of $\delta_{u}^{\left(\Omega\right)}$
and $\delta_{b}^{\left(\Omega\right)}$ agree with the $\tau$~approximation
calculation of \citet{Radler:2003gg} ($\delta_{u}^{\left(\Omega\right)}<0$,
$\delta_{b}^{\left(\Omega\right)}>0$, although there is not an exact
cancellation at $\bar{u}=\bar{b}$ as in \citet{Radler:2003gg}). 

The $\alpha$ effects arising through stratification and inhomogeneity also show broad agreement
with previous works. Because of the linearity of the expansion 
in $\nabla \ln \rho $, $U$ and $\Omega$, the density stratification contributes
very little to the coefficients, aside from directly through $\nabla W_{m}$ [Eq.~\eqref{eq:Wm deriv}]. This 
means $\chi_{\rho}$ generally appears together with the turbulent gradient 
$\chi_{\bar{u}}$.
The one exception to this is the ``turbulent diamagnetism'' term, $\gamma^{(0)}$, which 
interestingly depends only on the turbulence gradient, not the density gradient, due to a cancellation (this 
is in agreement with \citet{Kichatinov:1992vg}). Again our results 
without mean velocity
broadly agree with the $\tau$~approximation magnetic turbulence results given in \citet{Radler:2003gg};
for instance, the fact that $\left( \gamma^{0}\right)_{b}=-\left( \gamma^{0}\right)_{u}$ and 
the opposing signs of the rotational kinematic and magnetic
diagonal $\alpha$ effects $( \alpha^{\Omega}_{1})_{u,b}$, with 
$| ( \alpha^{\Omega}_{1})_{u}|>| ( \alpha^{\Omega}_{1})_{b}|$
(although we see a strong dependence of these parameters on Pm; see Sec.~\ref{sec:stratified}).

\section{Specific results for unstratified shear dynamos}\label{sec:Specific-results-for}

In this section we discuss the results pertinent to our primary motivation
for this work, the shear dynamo in a Cartesian box. As shown in Eq.~\eqref{eq:SC sa eqs},
in this geometry with a horizontal mean-field average, the number
of transport coefficients reduces significantly. We are particularly
interested in the sign of the $\eta_{yx}$ coefficient, which should
be most important for dynamo growth due to its coupling with the shear
{[}Eq.~\eqref{eq:SC growth}{]}. Here we outline the contribution
to $\eta_{yx}$ from velocity and magnetic fluctuations in the presence
of shear, both with and without rotation. This geometry is particularly relevant
for the central regions of accretion disks, where there is strong flow shear, stratification may be
subdominant, and there is no obvious source of helicity in either 
velocity or magnetic  fluctuations \cite{Lesur:2008cv}.

Utilizing Eq.~\eqref{eq: eta from beta} and the results in listed
in App.~\ref{sec:List-of-all transports coeffs}, one obtains after
some impressive cancellations 
\begin{equation}
(\eta_{yx})_{u}^{S}=\int d\omega\, dk\frac{32\pi k^{2}W_{u}(k,\omega)\,\omega^{2}\tilde{\eta}^{2}}{15\left(\tilde{\eta}^{2}+\omega^{2}\right)^{2}\left(\tilde{\nu}^{2}+\omega^{2}\right)},\label{eq:eta Su}
\end{equation}
\begin{align}
(\eta_{yx})_{b}^{S} & =\int d\omega\, dk\,8\pi k^{2}\rho^{-1}W_{b}\left(k,\omega\right)\left(\frac{4\omega^{4}}{15\left(\tilde{\nu}^{2}+\omega^{2}\right)^{3}}\right.\nonumber \\
 &\qquad\qquad- \frac{2\tilde{\eta}\tilde{\nu}^{3}+\tilde{\eta}^{2}\tilde{\nu}^{2}+2\omega^{2}\tilde{\eta}^{2}+3\omega^{4}}{15\left(\tilde{\eta}^{2}+\omega^{2}\right)\left(\tilde{\nu}^{2}+\omega^{2}\right)^{2}}\nonumber\\
 &\qquad\qquad+ \left.\frac{4\omega^{2}\tilde{\eta}\tilde{\nu}}{15\left(\tilde{\eta}^{2}+\omega^{2}\right)^{2}\left(\tilde{\nu}^{2}+\omega^{2}\right)}\right),\label{eq: eta Sb}
\end{align}
\begin{equation}
(\eta_{yx})_{u}^{\Omega}=-\int d\omega \,dk\frac{64\pi k^{2}W_{u}\left(k,\omega\right)\omega^{2}\tilde{\eta}^{2}}{15\left(\tilde{\eta}^{2}+\omega^{2}\right)^{2}\left(\tilde{\nu}^{2}+\omega^{2}\right)},\label{eq:eta Ou}
\end{equation}
\begin{equation}
(\eta_{yx})_{b}^{\Omega}=-\int d\omega \,dk\frac{8\pi k^{2}\rho^{-1}W_{b}\left(k,\omega\right)\left(\tilde{\nu}^{4}-12\omega^{2}\tilde{\nu}^{2}+3\omega^{4}\right)}{15\left(\tilde{\nu}^{2}+\omega^{2}\right)^{3}}.\label{eq:eta Ob}
\end{equation}
Here $\tilde{\nu}=\nu k^{2}$, $\tilde{\eta}=\eta k^{2}$, integration
over $\omega$ is from $-\infty$ to $\infty$ and over $k$ is from $0$
to $\infty$. We have defined each coefficient such that 
\begin{equation}
\eta_{yx}=S\left[(\eta_{yx})_{u}^{S}+(\eta_{yx})_{b}^{S}\right]+\Omega\left[(\eta_{yx})_{u}^{\Omega}+(\eta_{yx})_{b}^{\Omega}\right],
\end{equation}
to keep all signs consistent. Recall from Eq.~\eqref{eq:SC growth}
that with our definition of $S$, $\eta_{yx}S<0$ is required for
a growing dynamo (note that this is the reverse of \citetalias{Radler:2006hy}). For Keplerian 
rotation,  $\Omega=2S/3$, since vorticity and rotation are opposite (i.e., anticyclonic)
when $S$ and $\Omega$ have the same sign.
%%%%%%%%%%%%%%%%%%%%%%%%%
\begin{figure}
\begin{centering}
\includegraphics[width=0.8\linewidth]{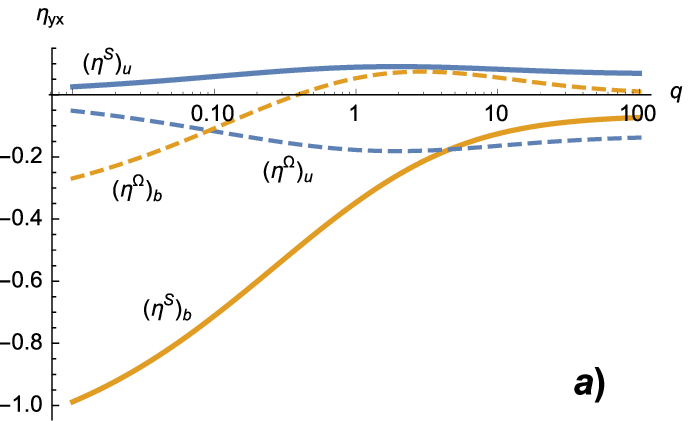}
\includegraphics[width=0.85\linewidth]{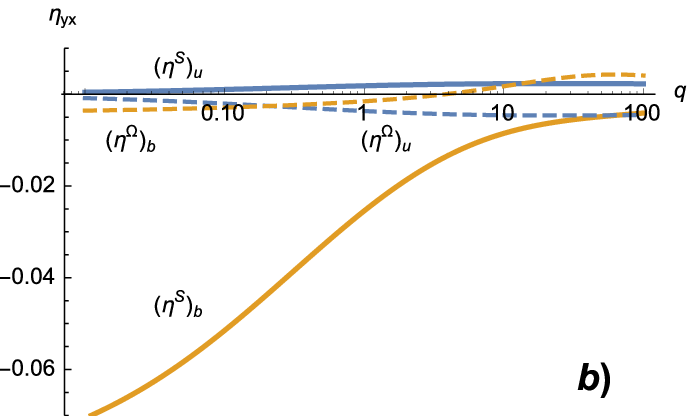}
\includegraphics[width=0.8\linewidth]{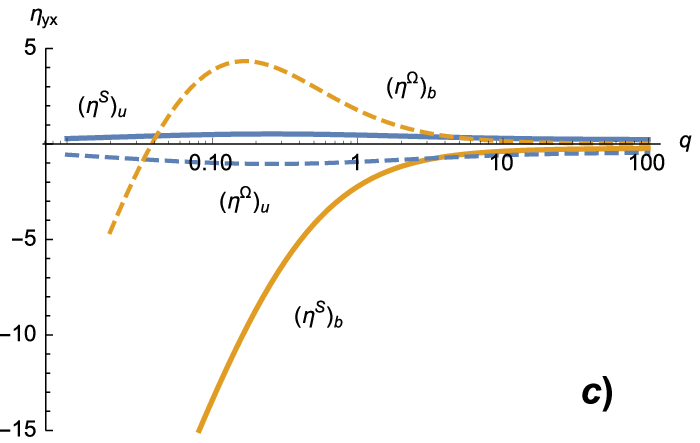}
\caption{Transport coefficients $(\eta_{yx})_{u}^{S}$ (solid, blue), $(\eta_{yx})_{u}^{\Omega}$ (dashed, blue), $(\eta_{yx})_{b}^{S}$ (solid, orange) and $(\eta_{yx})_{b}^{\Omega}$ (dashed, orange)
as a function of $q$ for (a) $\mathrm{Pm}=1$, (b) $\mathrm{Pm}=10$, and (c) $\mathrm{Pm}=1/10$.
Each coefficient has been calculated using
the form given in Eq.~\eqref{eq:gaussian W} for $W$, and normalized by $\left(\beta^{(0)}\right)_{u}$ with the
magnetic diffusion time, $\lambda_{c}^{2}/\eta$, held constant (equivalently $\tau_{c}=1/q$). (Note 
that this choice is necessary because the coefficients have different units, and
is chosen purely for plotting purposes, since it reduces the variation of coefficients
with $q$.)
\label{fig:Transport-coefficients-S} }
\end{centering}
\end{figure}
%%%%%%%%%%%%%%%%%%%%%%%%%%%%%%%%%%

Let us first examine the coefficients for a kinematic dynamo, i.e.,
with strong homogenous velocity fluctuations {[}the coefficients $(\eta_{yx})_{u}$,
Eqs.~\eqref{eq:eta Su} and \eqref{eq:eta Ou}{]}. Firstly, we note
that the contributions from $S$ and $\Omega$ have identical forms,
and that the integrands are positive definite
\footnote{We have found a different result for $(\eta_{yx})_{u}^{S}$
from \citetalias{Radler:2006hy}, in particular only obtaining the first
part of their Eq.~(D5), and are currently unsure from where this
discrepancy arises. We have one difference in the full transport coefficients
(in the $\beta^{(D)}$ term, see App.~\ref{sec:List-of-all transports coeffs}),
but this difference alone does not fix the discrepancy. In any case,
the main conclusion -- that $(\eta_{yx})_{u}^{S}$ has
the incorrect sign for dynamo action -- is unchanged. Our expressions
for $(\eta_{yx})_{u}^{\Omega}$ are identical.%
}, see Fig.~\ref{fig:Transport-coefficients-S}.
Thus, as is well known, we see that $(\eta_{yx})_{u}^{S}$,
the ``shear-current effect,'' has the incorrect sign for dynamo
action within this quasi-linear approximation. Although the basic
$\bm{\Omega}\times\bm{J}$ effect (also known as the Rädler effect)
is well known, the explicit calculation of transport coefficients
including shear and rotation seems to have been mostly ignored, although
there is much discussion in early literature on the subject (e.g.,
\citet{Krause:1980vr,Moffatt:1982ta}). Given the identical forms
of Eqs.~\eqref{eq:eta Su} and \eqref{eq:eta Ou}, we can immediately
write down the result 
\begin{equation}
(\eta_{yx})_{u}=\left(S-2\Omega\right)\Xi,
\end{equation}
where $\Xi$ is the (positive) integral in Eq.~\eqref{eq:eta Su}.
Thus, we find that the addition of Keplerian rotation ($\Omega=2S/3$)
(as relevant to turbulence in accretion disks for example), will change
the sign of $\eta_{yx}$ to slightly negative and a coherent dynamo
instability should be possible. Indeed, this is seen in our recent
simulation work \citep{LowRm}, where we observe increasing coherency
and a larger growth rate as the rotation is increased in the anticyclonic
direction. 

Turning to the coefficients for magnetic fluctuations we find the
interesting possibility of a magnetically driven dynamo. In particular,
as shown in App.~\eqref{sec:The-sign-of etab} and Fig.~\ref{fig:Transport-coefficients-S},
the coefficient $(\eta_{yx})_{b}$ is consistently negative and generally larger than 
the other contributions. This implies that
a dynamo can be excited by magnetic fluctuations, themselves presumably
arising from a small-scale dynamo process, or perhaps an MHD instability
of some sort. Since the small-scale dynamo is usually considered harmful
to mean fields \citep{Kulsrud:1992ej}, this is an interesting possibility -- a
 build of magnetic noise on small scales
may \emph{cause} a coherent large-scale dynamo to develop. The addition of
rotation renders the effect of magnetic fluctuations more complex,
and no simple result seems possible. In particular, the sign of the
$(\eta_{yx})_{b}^{\Omega}$ coefficient depends on the
parameters, and is generally negative for large $\nu$, $\eta$
and positive at lower dissipation, although smaller in magnitude than $(\eta_{yx})_{b}^{S}$
This change in sign is also seen in  quasi-linear calculations
\citep{LowRm}; however, given that
the quasi-linear approximation becomes less valid in this limit, it
would be unwise to draw any conclusions about the high-$\mathrm{Rm}$ limit from this behavior. 

Finally, we note the possible relevance of this dynamo to the 
central regions of accretion disks. In self-sustaining turbulence simulations
in this geometry, magnetic fluctuations are generally substantially 
stronger than velocity fluctuations. 
Such conditions seem ideal for excitation of a coherent  
dynamo driven by the magnetic shear-current effect. We note 
that cyclic behavior, as often observed in self-sustaining simulations \cite{Lesur:2008cv,Simon:2012dq},
seems to be quite generic in the nonlinear development of the magnetic 
shear current effect, and we have observed this in low-$\mathrm{Rm}$ simulations with a forced induction
equation \cite{LowRm}. In addition, it is worth noting 
that \citet{Lesur:2008cv} concluded that $\eta_{yx}$ was the primary 
dynamo driver from analysis of their numerical simulations.
While more work is obviously needed to explore this possibility 
in detail, it seems reasonable to conclude that the magnetic shear-current effect is playing
a fundamental role. 

\section{Specific results for stratified accretion disks}\label{sec:stratified}

In this section we briefly outline how our results apply to stratified sheared 
rotating turbulence. Our primary motivation is consideration of the upper and lower 
regions of accretion disks, where the turbulence is stratified in density and intensity by the 
vertical gravity, perpendicular to the velocity shear. Self-sustaining turbulence
simulations in this geometry (for instance with
shear-periodic boundary conditions in the radial direction) exhibit 
a very coherent dynamo, with
quasi-time-periodic behavior in $B_{y}$ and $B_{x}$ creating
a ``butterfly diagram'' \cite{Brandenburg:1995dc,Gressel:2010dj}.
Large-scale magnetic structures are seen to emanate from the central 
portion of the disk, migrating upwards into the lower density regions and
becoming more intense as they do so \cite{Simon:2012dq}. This migration behavior 
would be characteristic of a dynamo driven by $\alpha_{yy}$ above and below the mid-plane: as shown in 
Eq.~\eqref{eq:growth alpha}, growth of this type of ``$\alpha \omega$'' dynamo 
is always accompanied by dynamo waves since $\Gamma$ is complex. 
Note that a negative imaginary part of $\Gamma$ is required for upwards 
migration of mean-field structures with $\hat{\bm{g}}=\hat{\bm{z}}$. This
occurs for $a_{yy}<0,\:a_{xy}<0,\:(a_{yx}>0)$ \cite{Rudiger:2000wt}.

Utilizing Eq.~\eqref{eq:ayy def} with the results listed in App.~\ref{sec:List-of-all transports coeffs},
and setting $\mathrm{Pm}=1$ here for simplicity, one obtains,
\begin{equation}
(a_{yy})^{S}_{u}=8\pi\chi_{\rho \bar{u}}\int d\omega\, dk \frac{k^{2} W_{u}(k,u) \tilde{\nu} ^2 \left(5 \tilde{\nu} ^2+\omega ^2\right)}{15 \left(\tilde{\nu} ^2+\omega ^2\right)^3},\label{eq:ayySu}
\end{equation}
\begin{equation}
(a_{yy})^{S}_{b}=  -4\pi\chi_{ \bar{b}}\int d\omega\, dk\, \rho^{-1}W_{b}(k,u)k^{2} \frac{7 \tilde{\nu} ^4-4 \omega ^2 \tilde{\nu} ^2-3 \omega ^4}{15 \left(\tilde{\nu} ^2+\omega ^2\right)^3},\label{eq:ayySb}
\end{equation}
\begin{equation}
(a_{yy})^{\Omega}_{u}=  -64\pi\chi_{\rho \bar{u}}\int d\omega\, dk \frac{k^{2} W_{u}(k,u) \tilde{\nu} ^2 \left(\tilde{\nu} ^2+5 \omega ^2\right)}{15 \left(\tilde{\nu} ^2+\omega ^2\right)^3},\label{eq:ayyOu}
\end{equation}
\begin{equation}
(a_{yy})^{\Omega}_{b}=  -64\pi\chi_{ \bar{b}}\int d\omega\, dk \frac{ \rho^{-1}W_{b}(k,u)k^{2} \omega ^2 \left(\omega ^2-3 \tilde{\nu} ^2\right)}{15 \left(\tilde{\nu} ^2+\omega ^2\right)^3}.\label{eq:ayyOb}
\end{equation}
Finally, for the off-diagonal component, $\gamma^{(0)}=a_{xy}=-a_{yx}$, one has
\begin{equation}
(\gamma^{(0)})_{u}=4\pi\chi_{\bar{u}}\int d\omega \,dk\frac{k^{2}W_{u}(k,u)\tilde{\eta}}{3\left(\tilde{\eta}^{2}+\omega^{2}\right)},\label{eq:ayxSu}
\end{equation}
\begin{equation}
(\gamma^{(0)})_{b}=-4\pi\chi_{\bar{b}}\int d\omega \,dk\frac{k^{2}\rho^{-1}W_{b}(k,u)\tilde{\eta}}{3\left(\tilde{\eta}^{2}+\omega^{2}\right)}.\label{eq:ayxSb}
\end{equation}
Here we use the notation $\chi_{\rho \bar{u}}=|\nabla \ln (\rho  \bar{u}) |$, and again
signs are defined such that 
\begin{equation}
a_{yy}=S\left[(a_{yy})_{u}^{S}+(a_{yy})_{b}^{S}\right]+\Omega\left[(a_{yy})_{u}^{\Omega}+(a_{yy})_{b}^{\Omega}\right],\label{eq:full ayy}
\end{equation}
for anticyclonic rotation, e.g., Keplerian rotation is $\Omega=2/3S$.

\begin{figure}
\begin{centering}
\includegraphics[width=0.8\linewidth]{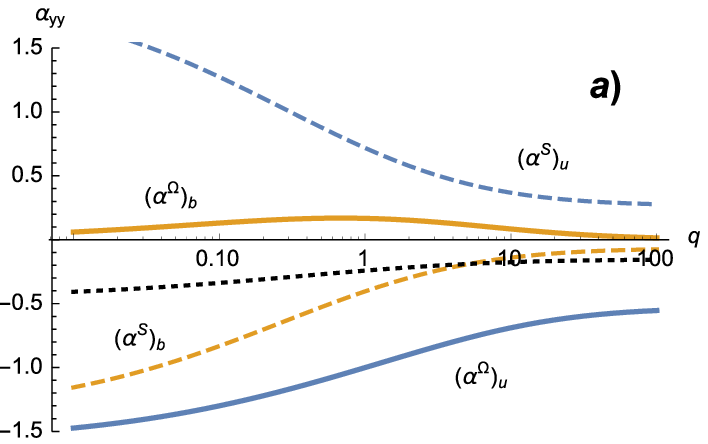}
\includegraphics[width=0.8\linewidth]{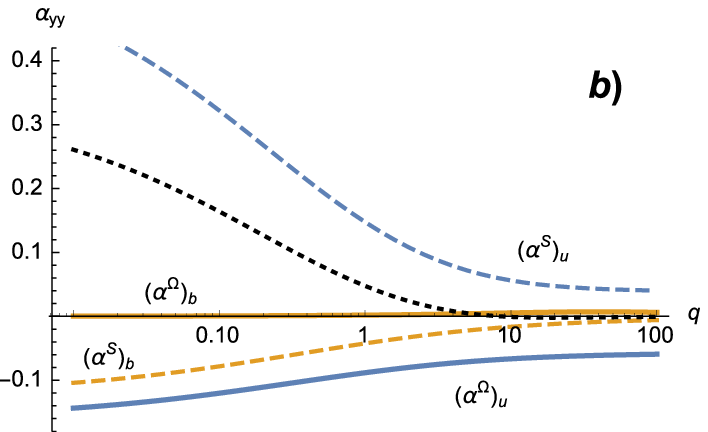}
\includegraphics[width=0.8\linewidth]{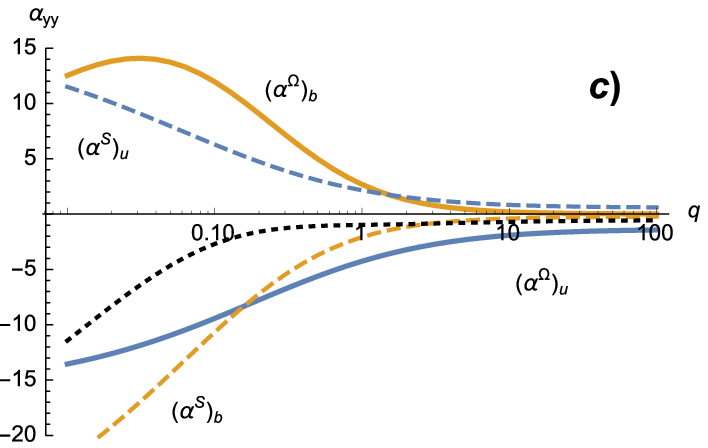}
\caption{Transport coefficients $(a_{yy})_{u}^{\Omega}$ (solid, blue), $(a_{yy})_{u}^{S}$ (dashed, blue), $(a_{yy})_{b}^{\Omega}$ (solid, orange) and $(a_{yy})_{b}^{S}$ (dashed, orange)
as a function of $q$ for (a) $\mathrm{Pm}=1$, (b) $\mathrm{Pm}=10$, and (c) $\mathrm{Pm}=1/10$.
Each coefficient has been calculated using
the form given in Eq.~\ref{eq:gaussian W} for $W$, and normalized by $\left(\beta^{(0)}\right)_{u}$ with the
magnetic diffusion time, $\lambda_{c}^{2}/\eta$, held constant (equivalently $\tau_{c}=1/q$).
The dotted (black) curve in each plot shows the total $a_{yy}$ with equal kinetic and magnetic turbulence levels
for Keplerian rotation, $\Omega=2/3S$ [Eq.~\eqref{eq:full ayy}],  
to illustrate the variability in these predictions. 
\label{fig:ayy coefficients} }
\end{centering}
\end{figure}

It is first worth noting the sign of each coefficient given in Eqs.~\eqref{eq:ayySu}-\eqref{eq:ayxSb}. With
$\chi_{\rho\bar{u}},\,\chi_{\bar{b}}>0$ it can be shown easily from the above expressions that
\begin{equation}
(a_{yy})^{S}_{u}>0,\enspace(a_{yy})^{S}_{b}<0,\enspace(a_{yy})^{\Omega}_{u}<0,\enspace(a_{yy})^{\Omega}_{b}>0. \label{eq:ayy relations}
\end{equation}
(Note that for the $b$ components, it is necessary to integrate by parts over $\omega$, 
see App.~\ref{sec:The-sign-of etab}). The relations in Eq.~\eqref{eq:ayy relations} 
appear to also hold for $\mathrm{Pm}\neq1$ (although we have a proof of this only for the $\Omega$ coefficients). 
This consistent difference in sign between contributions is rather  inconvenient for the application of 
SOCA results to stratified accretion disks. Since one expects $\chi_{\rho \bar{u}}<0,\:\chi_{ \bar{b}}<0$ 
(although possibly $\chi_{ \bar{u}}>0 $) \cite{Gressel:2010dj, Bodo:2014fx}, 
we are left with the situation where not only do the $\alpha$ effects due to $u$ and $b$ partially cancel, but 
also those due to rotation and velocity shear! What's more, as shown 
in Fig.~\ref{fig:ayy coefficients}, the relative contribution of each depends strongly on Pm. 
%%%%%%%%%%%%%%%%%%%%%%%%%
%%%%%%%%%%%%%%%%%%%%%%%%%%%%%%%%%%
In particular, we see a dominance of $(a_{yy})_{u}$ over $(a_{yy})_{b}$ for $\mathrm{Pm}\gtrsim1$, 
but this can reverse at low $\mathrm{Pm}$. Similarly, the relative contributions 
due to velocity shear and rotation for the magnetic effect vary substantially with $\mathrm{Pm}$,
although the effect of shear seems generally more substantial. While the ratio of
kinematic shear and rotation contributions may be somewhat more robust, 
the two are roughly equal in magnitude, 
$\enspace(a_{yy})^{S}_{u}\sim-(a_{yy})^{\Omega}_{u}$,
and will approximately cancel for Keplerian rotation. 
Finally, it is worth noting that to complement these uncertainties,
the signs of $\gamma^{(0)}$ seem to predict the \emph{opposite} field 
migration pattern to the upwards transport seen in simulation. In particular, 
for $\chi_{ \bar{b}}<0,\:\chi_{ \bar{u}}>0 $, the kinematic and magnetic contributions
both enforce $\gamma^{(0)}>0$, leading to $\mathrm{Im}\, \Gamma >0$.
However, in our use of the anelastic approximation, buoyancy effects
are not included and these would be expected to change this aspect of the calculation
substantially  \citep{Kichatinov:1993ti,Moss:1999ua,Rudiger:2000wt}, potentially through
large-scale instability \citep{Rozyczka:1995uw}.

Where does this leave us for understanding the dynamo in stratified accretion disks? 
We see that aside from perhaps the transport term $\gamma^{(0)}$,
claims that SOCA predictions are  \emph{incorrect} for the stratified regions
of accretion disks are unfounded. More accurately, one could say 
that SOCA predictions themselves are completely inconclusive, even in the kinematic regime, 
since each contribution -- kinematic, magnetic, rotation, and velocity shear -- has a tendency to cancel its 
partner. Such uncertainty seems at odds with the robust dynamo ``butterfly diagram''
seen across a wide variety accretion disk of simulations. 

Of course, one possibility is that the  SOCA calculation carried out here, keeping
only the linear contributions due to $\Omega$, $S$ and stratification, is not
up to the task of calculating these coefficients, and in reality there \emph{is}
a robust $\alpha$ effect. For instance, in \citet{Rudiger:2000wt}, the 
authors find that $\alpha_{yy}$ has the correct sign ($\alpha_{yy}<0$) for 
magnetic fluctuations in a compressible turbulence model for Keplerian shear and moderate $\mathrm{Pm}$
(this is the  sign opposite to Eq.~\eqref{eq:ayySb} but since their effect vanishes in 
the incompressible limit, one should have no reason to expect agreement).
Similarly, the calculations presented in \citet{Donnelly:2013tu} 
go well beyond the accuracy of SOCA for the specific case of Keplerian shear
through non-perturbative inclusion of several extra physical effects; however, 
it is unclear from their (rather complicated) expressions whether the theory 
predicts a specific sign for $\alpha_{yy}$. 
While certainly feasible, it would seem a little bizarre
that a behavior that appears so robustly in simulation could 
show so much variability across different calculation methods or
rely on nonlinear behavior of transport coefficients with $\Omega$,
$S$, or the stratification. A variety of other possibilities might
be imaginable, for instance a dynamo driven primarily by the magnetic shear-current
effect up to relatively far from the mid-plane (Sec.~\ref{sec:Specific-results-for}), with upwards transport 
above this caused by large-scale buoyant instability (not included here due to the anelastic approximation).
Another possibility could  be 
that upwards field transport  is
caused by a small-scale magnetic helicity flux \cite{Vishniac:2001wo,Subramanian:2004gm} 
from  the central shear-current dynamo,
which would create a (helical) magnetic $\alpha$ effect. Such an process could look
rather similar to a more standard $\alpha$ effect, although the 
basic cause of the dynamo would be entirely different \cite{Gressel:2010dj}. 
Note that magnetic helicity fluxes have been found to be
playing a significant role in unstratified global MRI turbulence \cite{Ebrahimi:2014jt}, providing 
some indication that such a process could be important. It is also worth noting
that spatial variation in transport coefficients and quenching can lead to 
some interesting possibilities for dynamo action \cite{Parker:1993kd,Tobias:1996eu}, and 
similar effects may prove important at the boundary between the stratified and
unstratified regions of disks. 
Overall however, it seems that the underlying cause for the 
``butterfly diagram'' in stratified disks remains unclear and more 
work will be needed to arrive at robust mean-field models of the process.

%To avoid this variability in the sign of $\alpha$, we propose another possibility -- that 
%the migration pattern in stratified disks does
%not stem from a standard $\alpha$ effect at all. Instead, the primary driver
%for the dynamo could arise in the central regions of the disk, from the magnetic
%shear-current effect discussed in Sec.~\ref{sec:Specific-results-for}. As this dynamo causes 
%large scale field growth, a small-scale current helicity develops, consistent with the 
%ideas of helicity conservation \cite{Field:2002fn,Subramanian:2004gm}.
%Due to the inhomogeneity of the turbulence, such a small-scale 
%helicity would be transported upwards ,Ebrahimi:2014jt}, 
%generating large scale fields due to the magnetic $\alpha$ effect. 
%Presumably a phase lag between small-scale current helicity and large
%scale field generation would cause migration of the large scale structures, 
%perhaps aided by buoyancy effects.
%Such a scenario is in line with proposals that accretion disks 
%show two separate dynamos . 
%We note that the effect would presumably look like a standard $\alpha$ effect
%in simulations (for instance in the distribution of current helicity between the upper and
%lower regions

\section{Discussion and conclusions}

In this work we have theoretically studied the dynamo in systems with
mean velocity gradients, rotation, net helicity, and stratification, using perturbative calculations within
the second-order correlation approximation. In addition to the standard
kinematic dynamo, we have considered the possibility of a dynamo driven
by small-scale \emph{magnetic fluctuations,} as might arise from the
small-scale dynamo or an instability.  Our
main finding is that an off-diagonal resistivity coupled to the shear
can cause a dynamo instability in the presence of magnetic fluctuations.
This effect -- the magnetic analogue of the ``shear-current effect''
\citep{Rogachevskii:2003gg,Rogachevskii:2004cx} -- raises the interesting
possibility of the small-scale dynamo \emph{enhancing} the growth
of a large-scale field. In some sense, this possibility is the reverse
of large-scale quenching \cite{Kulsrud:1992ej,Blackman:2002fe}; rather than
the small-scale magnetic fluctuations inhibiting the large-scale field
growth, they could actively aid field generation, with large-scale
growth eventually halting due to nonlinear changes to the transport
coefficients, possibly influenced by secondary quenching effects \cite{Rogachevskii:2006hy}.

Importantly, our prediction that the magnetic shear-current effect
is able to excite a dynamo agrees with other transport coefficient calculation methods and simulations.
In particular, the $\tau$~approximation predicts the linear magnetic
effect to be much stronger than the kinematic effect (see Fig.~3
of \citet{Rogachevskii:2004cx}), just as was found in this work using SOCA (Fig.~\ref{fig:Transport-coefficients-S}).  
In addition, agreement is found with
 quasi-linear calculations \citep{LowRm} (the magnetic version
of the calculations in \citet{SINGH:2011ho}), as well as perturbative 
inhomogenous shearing wave calculations \cite{Lesur:2008fn}. This suggests that
the effect may be more robust than the kinematic shear-current effect
and/or have less dependence on Reynolds numbers.

The work presented in this manuscript was primarily motivated by gaining
improved understanding of the fundamental dynamo mechanisms in accretion
disks. Consistent with the idea that two dynamo 
mechanisms might operate in disks \cite{Blackman:2004hf}, 
their inner regions seem well suited to 
be explained by the magnetic shear current effect \cite{Lesur:2008cv} -- magnetic fluctuations are 
generally stronger than kinetic fluctuations, rotation has the correct
sign to enhance the kinematic dynamo, and the turbulence 
is essentially unstratified and nonhelical.
Concurrent nonlinear direct numerical simulations
of unstratified shear dynamos in Cartesian boxes \citep{LowRm,HighRm} have confirmed 
all results discussed in Sec.~\ref{sec:Specific-results-for}
for the low-$\mathrm{Rm}$ regime \cite{Yousef:2008ie,LowRm}. Firstly, we see a qualitative
change in the kinematic dynamo with the addition of rotation, due to
the change in sign of the $\eta_{yx}$ transport coefficient \citep{LowRm}. 
Secondly, we observe the magnetically driven shear-current effect, both through
direct driving of the induction equation \citep{LowRm}, and 
at higher magnetic Reynolds number
where  magnetic fluctuations arise self-consistently though excitation of a small-scale dynamo \citep{HighRm}. 
The nonlinear saturation of these magnetically driven large-scale dynamos
exhibits a pleasing resemblance to self-sustaining unstratified accretion disk turbulence 
simulations, with quasi-cyclic behavior of the large-scale $B_{y}$ field. 

Less clear have been our findings regarding the $\alpha$ effect, as relevant to the 
stratified regions of accretion disks. In particular, we find that $\alpha$ coefficients arising from
rotation and shear, and those arising from kinetic and magnetic fluctuations, 
are each of opposite signs for anticyclonic rotation ($\bm{\Omega}$ and $\nabla\times \bm{U}$ antiparallel), and
thus would tend to cancel. Furthermore,  
predictions about which of these terms dominate (thus determining
the sign of the total $\alpha$ effect), depend strongly on the
magnetic Prandtl number and the relative levels of kinetic and magnetic turbulence. 
We thus conclude that perturbative SOCA calculations \emph{give no useful predictions} regarding
the primary driver of the so-called ``butterfly diagram'' pattern of 
large-scale field evolution seen in self-sustaining stratified
accretion disk simulations. Whether this is simply due to 
the inaccuracies of SOCA or there is some other more exotic effect
operating (e.g., a helicity flux \cite{Ebrahimi:2014jt}), remains to be seen.

%Of course, the formal applicability of the second-order correlation
%approximation is limited to very low $\mathrm{Rm}$ and Re or low St. As such, we see the
%utility in such calculations as giving a useful indication of the
%expected forms of transport coefficients, in particular their sign,
%rather than detailed quantitative prediction. Nonetheless, SOCA has
%often been seen to be surprisingly accurate \cite{Brandenburg:2005kla}
%even outside of its formal validity range, and the agreement with
%other calculation methods and nonlinear simulation gives us some confidence
%that this should be the case for the magnetic shear-current effect studied here also. 
%One interesting avenue for future analytic work could be an exploration of vorticity dynamos
%in these shearing rotating systems, in particular how self-generated
%flows could interact with the small and large scale dynamo fields.
%Such a study could be done in essentially the same framework used
%throughout this work. 

%\bibliographystyle{apsrev4-1}
%\bibliography{fullbibPRE}


\appendix

\section{Equations for $\bm{u}^{(0)},\,\bm{u}^{(1)},\,\bm{b}^{(0)},\,\bm{b}^{(1)}$
in Fourier space}\label{app:Fourier space}

Here we give the set of perturbation equations for $\bm{u}$ and $\bm{b}$
in Fourier space, as result from the Fourier transform of Eq.~\eqref{eq:perturbation expansion}
The method is outlined in \citetalias{Radler:2006hy}, so we give very
little detail here. Since we assume $U_{i}\left(\bm{x}\right)=U_{ij}x_{j}$, $\rho=\rho_{0}+\chi_{\rho}\hat{g}_{i}x_{i}$,
and $B_{i}\left(\bm{x}\right)=B_{i}+B_{ij}x_{j}$ the Fourier transforms
can be carried out exactly using $\widehat{x_{k}\partial_{l}b_{j}}=-\delta_{lk}\hat{b}_{j}-k_{l}\partial_{k_{k}}\hat{b}_{j}$
(where $\hat{\cdot}$ denotes the Fourier transform). We have also neglected 
products of $\chi_{\rho}$ with $B_{ij}$. In the momentum
equations, the projection operator $\delta_{ij}-k_{i}k_{j}/k^{2}$
is applied so as to remove the pressure. 

Defining, as in \citetalias{Radler:2006hy}, 
\begin{equation}
N_{\nu}=\frac{1}{i\omega-\nu k^{2}},\:\: E_{\eta}=\frac{1}{i\omega-\eta k^{2}},
\end{equation}
the Fourier space equations are as follows,

\begin{align}
m_{i}^{(0)}=&N_{\nu}  \left[-U_{il}m_{0l}+U_{lk}k_{l}\frac{\partial m_{0i}}{\partial k_{k}}+2\frac{k_{i}k_{j}}{k^{2}}m_{0l}U_{jl}-i\nu k_{r}\chi_{\rho}\hat{g}_{r} m_{0i}\right.\nonumber \\
 & +i\nu \frac{k_{i}k_{j}k_{r}}{k^{2}}\chi_{\rho}\hat{g}_{r} m_{0j}+2\frac{k_{r}\Omega_{r}}{k^{2}}\varepsilon_{ijk}m_{0j}k_{k}+ik_{r}B_{r}b_{0i}\nonumber\\
 & -ik_{r}B_{r}\frac{k_{i}k_{j}}{k^{2}}b_{0j}+ \left.B_{il}b_{0l}-B_{lk}k_{l}\frac{\partial b_{0i}}{\partial k_{k}}-2\frac{k_{i}k_{j}}{k^{2}}b_{0l}B_{jl}\right],\label{eq:u0FS}
\end{align}
\begin{align}
m_{i}^{(1)}=&N_{\nu}  \left[-U_{il}m_{l}^{(0)}+U_{lk}k_{l}\frac{\partial m_{i}^{(0)}}{\partial k_{k}}+2\frac{k_{i}k_{j}}{k^{2}}m_{l}^{(0)}U_{jl}-i\nu k_{r}\chi_{\rho}\hat{g}_{r} m_{i}^{(0)}\right.\nonumber \\
 & +i\nu \frac{k_{i}k_{j}k_{r}}{k^{2}}\chi_{\rho}\hat{g}_{r} m_{j}^{(0)}+2\frac{k_{r}\Omega_{r}}{k^{2}}\varepsilon_{ijk}m_{j}^{(0)}k_{k}+ik_{r}B_{r}b_{i}^{(0)}\nonumber\\
 & -ik_{r}B_{r}\frac{k_{i}k_{j}}{k^{2}}b_{j}^{(0)}+ \left.B_{il}b_{l}^{(0)}-B_{lk}k_{l}\frac{\partial b_{i}^{(0)}}{\partial k_{k}}-2\frac{k_{i}k_{j}}{k^{2}}b_{l}^{(0)}B_{jl}\right],\label{eq:u1FS}
\end{align}
\begin{align}
b_{i}^{(0)}=&E_{\eta}  \left[\rho_{0}^{-1}\left(ik_{r}B_{r}m_{0i}-B_{ij}m_{0j}-B_{jk}k_{j}\frac{\partial m_{0i}}{\partial k_{k}}+B_{i}\chi_{\rho}\hat{g}_{r} m_{0r}\right.\right.\nonumber \\
 &\left.\left. +\chi_{\rho}\hat{g}_{r} B_{j}k_{j}\frac{\partial m_{0i}}{\partial k_{r}}\right)+U_{ij}b_{0j}+U_{jk}k_{j}\frac{\partial b_{0i}}{\partial k_{k}}\right],\label{eq:b0FS}
\end{align}
\begin{align}
b_{i}^{(1)}=&E_{\eta}  \left[\rho_{0}^{-1}\left(ik_{r}B_{r}m_{i}^{(0)}-B_{ij}m_{j}^{(0)}-B_{jk}k_{j}\frac{\partial m_{i}^{(0)}}{\partial k_{k}}+B_{i}\chi_{\rho}\hat{g}_{r} m_{r}^{(0)}\right.\right.\nonumber \\
 &\left.\left. +\chi_{\rho}\hat{g}_{r} B_{j}k_{j}\frac{\partial m_{i}^{(0)}}{\partial k_{r}}\right)+U_{ij}b_{j}^{(0)}+U_{jk}k_{j}\frac{\partial b_{i}^{(0)}}{\partial k_{k}}\right],\label{eq:b1FS}
\end{align}
Here $m_{0i}$, $b_{0i}$ etc.~refer to the Fourier space variables for
simplicity of notation. As a first step in the calculation, Eqs.~\eqref{eq:u0FS}
and \eqref{eq:b0FS} are inserted into Eqs.~\eqref{eq:u1FS} and
\eqref{eq:b1FS} and expanded, neglecting those terms that contain
$U_{ij}U_{rs},$ $U_{ij}\Omega_{r},$ $\Omega_{i}\Omega_{j},$ $U_{ij}\,\chi_{\rho},$ $\Omega\,\chi_{\rho},$
$\chi_{\rho}^{2},$ $B_{i}B_{j}$,
$B_{i}B_{ij}$ and $B_{ij}B_{rs}$ as higher order in this perturbation
expansion.

\section{List of all transport coefficients}\label{sec:List-of-all transports coeffs}

In this Appendix we list all transport coefficients $\alpha^{(0)}$
$\beta^{(0)},$ $\delta^{(\Omega)},\dots$ in the form of integrals
over the isotropic velocity and magnetic correlation functions,
$W_{u}\left(\bm{R},k,\omega\right),$ $H_{u}\left(k,\omega\right),$$W_{b}\left(bm{R},k,\omega\right),$
$H_{b}\left(k,\omega\right)$. This parallels Appendix B in \citetalias{Radler:2006hy}
and there is some overlap; however, for completeness we list all coefficients. 

Analogous to the relations in Sec.~\ref{sec:Specific-results-for} for the Cartesian case and
\citetalias{Radler:2006hy}, we list here the coefficient of $4\pi k^{2}W_{u,b}$ or $4\pi k^{2}H_{u,b}$
in the integrand of each transport coefficient; that is $\tilde{\alpha}^{(\cdot)}_{H}$, $\tilde{\alpha}^{(\cdot)}$ or 
$\tilde{\beta}^{(\cdot)}$ in 
\begin{gather}
\left(\alpha_{H}^{(\cdot)}\right)_{u,b}=4\pi\int dk\,d\omega\, k^{2}\tilde{\alpha}^{(\cdot)}_{H}\left(k,\omega\right)H_{u,b}\left(k,\omega\right),\nonumber \\
\left(\alpha^{(\cdot)} \right)_{u,b}=4\pi\int dk\,d\omega\, k^{2}\tilde{\alpha}^{(\cdot)}\left(k,\omega\right)W_{u,b}\left(k,\omega\right),\nonumber \\
\left(\beta^{(\cdot)}\right)_{u,b}=4\pi\int dk\,d\omega\,k^{2}\tilde{\beta}^{(\cdot)}\left(k,\omega\right)W_{u,b}\left(k,\omega\right).
\end{gather}
We use the notation $\tilde{\eta}=k^{2}\eta$,
$\tilde{\nu}=k^{2}\nu$, and $ \nabla \ln a \equiv \chi_{a}\,\hat{\bm{g}} $ (e.g., $ \nabla \ln \rho+ \nabla \ln \bar{u} = \chi_{\rho\bar{u}}\,\hat{\bm{g}}$)

\subsection{Nonhelical $\alpha$ coefficients}

\begin{equation}
\left(\gamma^{(0)}\right)_{u}=\frac{\chi_{\bar{u}}\tilde{\eta}}{6\left(\tilde{\eta}^{2}+\omega^{2}\right)},
\end{equation}
%%%%%%%%%%%%%%%%%%%
\begin{equation}
\left(\gamma^{(0)}\right)_{b}=-\frac{\chi_{\bar{b}}\tilde{\nu}}{6\rho \left(\tilde{\nu}^{2}+\omega^{2}\right)},
\end{equation}
%%%%%%%%%%%%%%%%%%%
\begin{equation}
\left(\gamma^{(\Omega)}\right)_{u}=-\frac{\chi_{\rho\bar{u}}\omega^{2}}{3\left(\tilde{\eta}^{2}+\omega^{2}\right)\left(\tilde{\nu}^{2}+\omega^{2}\right)},
\end{equation}
%%%%%%%%%%%%%%%%%%%
\begin{equation}
\left(\gamma^{(\Omega)}\right)_{b}=\frac{\chi_{\bar{b}}(\omega^{2}-\tilde{\nu}^{2})}{6\rho\left(\tilde{\nu}^{2}+\omega^{2}\right)^{2}},
\end{equation}
%%%%%%%%%%%%%%%%%%%
\begin{align}
&\left(\alpha_{1}^{(\Omega)}\right)_{u}=\nonumber \\
&\quad\frac{4\chi_{\rho\bar{u}}\tilde{\eta}\left[2\omega^{2}\tilde{\eta}\left(\tilde{\nu}^{2}+\omega^{2}\right)+\tilde{\eta}^{2}\left(3\omega^{2}\tilde{\nu}+\tilde{\nu}^{3}\right)+\omega^{2}\tilde{\nu}\left(\tilde{\nu}^{2}+3\omega^{2}\right)\right]}{15\left(\tilde{\eta}^{2}+\omega^{2}\right)^{2}\left(\tilde{\nu}^{2}+\omega^{2}\right)^{2}},
\end{align}
%%%%%%%%%%%%%%%%%%%
\begin{equation}
\left(\alpha_{1}^{(\Omega)}\right)_{b}=\frac{4\chi_{\bar{b}}\omega^{2}\left(\omega^{2}-3\tilde{\nu}^{2}\right)}{15\rho\left(\tilde{\nu}^{2}+\omega^{2}\right)^{3}},
\end{equation}
%%%%%%%%%%%%%%%%%%%
\begin{align}
\left(\alpha_{2}^{(\Omega)}\right)_{u} & =\frac{\chi_{\rho\bar{u}}}{15}\left[2\omega^{2}\tilde{\eta}\tilde{\nu}\left(\omega^{2}-3\tilde{\nu}^{2}\right)+3\omega^{2}\tilde{\eta}^{2}\left(\tilde{\nu}^{2}+\omega^{2}\right) \right.\nonumber\\ 
&\left.+2\tilde{\eta}^{3}\tilde{\nu}\left(\omega^{2}-3\tilde{\nu}^{2}\right) -5\omega^{4}\left(\tilde{\nu}^{2}+\omega^{2}\right)\right]\nonumber\\
 &\times\left(\tilde{\eta}^{2}+\omega^{2}\right)^{-2}\left(\tilde{\nu}^{2}+\omega^{2}\right)^{-2},
\end{align}
%%%%%%%%%%%%%%%%%%%
\begin{equation}
\left(\alpha_{2}^{(\Omega)}\right)_{b}=\frac{\chi_{\bar{b}}(3\omega^{4}-24\omega^{2}\tilde{\nu}^{2}+5\tilde{\nu}^{4})}{30\rho\left(\tilde{\nu}^{2}+\omega^{2}\right)^{3}},
\end{equation}
%%%%%%%%%%%%%%%%%%%
\begin{align}
\left(\alpha_{1}^{(W)}\right)_{u} & =\frac{\chi_{\rho\bar{u}}}{120}\left[4\tilde{\eta}^{5}\left(11\omega^{2}\tilde{\nu}+5\tilde{\nu}^{3}\right)+4\tilde{\eta}\left(11\omega^{6}\tilde{\nu}+5\omega^{4}\tilde{\nu}^{3}\right)\right.\nonumber\\+
 & 8\tilde{\eta}^{3}\left(11\omega^{4}\tilde{\nu}+5\omega^{2}\tilde{\nu}^{3}\right)+\tilde{\eta}^{4}\left(12\omega^{2}\tilde{\nu}^{2}-\tilde{\nu}^{4}+13\omega^{4}\right)\nonumber\\-
 & \left.4\tilde{\eta}^{2}\left(5\omega^{4}\tilde{\nu}^{2}+3\omega^{2}\tilde{\nu}^{4}+2\omega^{6}\right)+5\omega^{4}\tilde{\nu}^{4}-5\omega^{8}\right]\nonumber\\ \times
 & \,\left(\tilde{\eta}^{2}+\omega^{2}\right)^{-3}\left(\tilde{\nu}^{2}+\omega^{2}\right)^{-2},
\end{align}
%%%%%%%%%%%%%%%%%%%
\begin{align}
\left(\alpha_{1}^{(W)}\right)_{b} & =\frac{\chi_{\bar{b}}}{120}\left[4\omega^{2}\tilde{\eta}\tilde{\nu}\left(\tilde{\nu}^{2}+\omega^{2}\right)^{2}  +\tilde{\eta}^{4}\left(\tilde{\nu}^{4}-36\omega^{2}\tilde{\nu}^{2}+11\omega^{4}\right) \right.\nonumber\\
 & -4\tilde{\eta}^{3}\tilde{\nu}\left(\tilde{\nu}^{2}+\omega^{2}\right)^{2}+4\tilde{\eta}^{2}\left(-11\omega^{4}\tilde{\nu}^{2}+5\omega^{2}\tilde{\nu}^{4}+8\omega^{6}\right)
 \nonumber\\-
 & \left.8\omega^{6}\tilde{\nu}^{2}+19\omega^{4}\tilde{\nu}^{4}+21\omega^{8}\right]\left(\tilde{\eta}^{2}+\omega^{2}\right)^{-2}\left(\tilde{\nu}^{2}+\omega^{2}\right)^{-3}\rho^{-1},
\end{align}
%%%%%%%%%%%%%%%%%%%
\begin{align}
\left(\alpha_{2}^{(W)}\right)_{u} & =\frac{\chi_{\rho\bar{u}}}{240}\left[-4\tilde{\eta}^{5}\left(3\omega^{2}\tilde{\nu}+5\tilde{\nu}^{3}\right)-4\tilde{\eta}\left(3\omega^{6}\tilde{\nu}+5\omega^{4}\tilde{\nu}^{3}\right)\right. \nonumber\\+
 & \tilde{\eta}^{4}\left(44\omega^{2}\tilde{\nu}^{2}+13\tilde{\nu}^{4}+31\omega^{4}\right)-8\tilde{\eta}^{3}\left(3\omega^{4}\tilde{\nu}+5\omega^{2}\tilde{\nu}^{3}\right)\nonumber\\
 & -28\tilde{\eta}^{2}\left(5\omega^{4}\tilde{\nu}^{2}+3\omega^{2}\tilde{\nu}^{4}+2\omega^{6}\right) \nonumber\\ &\left.+5\left(8\omega^{6}\tilde{\nu}^{2}+3\omega^{4}\tilde{\nu}^{4}+5\omega^{8}\right)\right]\nonumber\\ 
 & \qquad\times\left(\tilde{\eta}^{2}+\omega^{2}\right)^{-3}\left(\tilde{\nu}^{2}+\omega^{2}\right)^{-2},
\end{align}
%%%%%%%%%%%%%%%%%%%
\begin{align}
\left(\alpha_{2}^{(W)}\right)_{b} & =\frac{\chi_{\bar{b}}}{240}\left[28\omega^{2}\tilde{\eta}\tilde{\nu}\left(\tilde{\nu}^{2}+\omega^{2}\right)^{2}-28\tilde{\eta}^{3}\tilde{\nu}\left(\tilde{\nu}^{2}+\omega^{2}\right)^{2}\right.\nonumber\\
 & +\tilde{\eta}^{4}\left(-12\omega^{2}\tilde{\nu}^{2}+7\tilde{\nu}^{4}-3\omega^{4}\right) \nonumber \\ &-4\tilde{\eta}^{2}\left(17\omega^{4}\tilde{\nu}^{2}-5\omega^{2}\tilde{\nu}^{4}+14\omega^{6}\right)\nonumber\\ -
56 &  \left.\omega^{6}\tilde{\nu}^{2}+13\omega^{4}\tilde{\nu}^{4}-53\omega^{8}\right]\left(\tilde{\eta}^{2}+\omega^{2}\right)^{-2}\left(\tilde{\nu}^{2}+\omega^{2}\right)^{-3}\rho^{-1},
\end{align}
%%%%%%%%%%%%%%%%%%%
\begin{align}
\left(\alpha^{(D)}\right)_{u} & =\frac{\chi_{\rho\bar{u}}}{120}\left[12\omega^{2}\tilde{\eta}^{2}\tilde{\nu}^{2}\left(\tilde{\nu}^{2}+\omega^{2}\right)+12\tilde{\eta}^{5}\tilde{\nu}\left(\omega^{2}-\tilde{\nu}^{2}\right)\right. \nonumber\\+
  \,4\omega^{4}&\tilde{\eta}\tilde{\nu}\left(\tilde{\nu}^{2}+7\omega^{2}\right)+8\tilde{\eta}^{3}\left(5\omega^{4}\tilde{\nu}-\omega^{2}\tilde{\nu}^{3}\right)+5\omega^{4}\tilde{\nu}^{4}-5\omega^{8} \nonumber\\-
\tilde{\eta}^{4} &  \left.\left(20\omega^{2}\tilde{\nu}^{2}+9\tilde{\nu}^{4}+11\omega^{4}\right)\right]\left(\tilde{\eta}^{2}+\omega^{2}\right)^{-3}\left(\tilde{\nu}^{2}+\omega^{2}\right)^{-2},
\end{align}
%%%%%%%%%%%%%%%%%%%
\begin{align}
\left(\alpha^{(D)}\right)_{b} & =\frac{\chi_{\bar{b}}}{120}\left[-4\omega^{2}\tilde{\eta}\tilde{\nu}\left(6\omega^{2}\tilde{\nu}^{2}+\tilde{\nu}^{4}+5\omega^{4}\right)\right.\nonumber\\ 
& -\tilde{\eta}^{4}\left(12\omega^{2}\tilde{\nu}^{2}-5\tilde{\nu}^{4}+\omega^{4}\right) \nonumber\\ & +
  4\tilde{\eta}^{3}\tilde{\nu}\left(6\omega^{2}\tilde{\nu}^{2}+5\tilde{\nu}^{4}+\omega^{4}\right)\nonumber\\ &+4\tilde{\eta}^{2}\left(-3\omega^{4}\tilde{\nu}^{2}+3\omega^{2}\tilde{\nu}^{4}+2\omega^{6}\right)\nonumber\\+
 & \left.7\omega^{4}\tilde{\nu}^{4}+9\omega^{8}\right]\left(\tilde{\eta}^{2}+\omega^{2}\right)^{-2}\left(\tilde{\nu}^{2}+\omega^{2}\right)^{-3}\rho^{-1},
\end{align}
%%%%%%%%%%%%%%%%%%%
\begin{align}
\left(\gamma^{(W)}\right)_{u}&=-\frac{\chi_{\rho\bar{u}}}{120}\left[-8\omega^{6}\tilde{\eta}\tilde{\nu}-16\omega^{4}\tilde{\eta}^{3}\tilde{\nu}-8\omega^{2}\tilde{\eta}^{5}\tilde{\nu}\right.\nonumber\\
-\tilde{\eta}^{4}&\left(8\omega^{2}\tilde{\nu}^{2}+\tilde{\nu}^{4}+7\omega^{4}\right)-4\tilde{\eta}^{2}\left(7\omega^{4}\tilde{\nu}^{2}+3\omega^{2}\tilde{\nu}^{4}+4\omega^{6}\right)\nonumber\\
+&\left.12\omega^{6}\tilde{\nu}^{2}+5\omega^{4}\tilde{\nu}^{4}+7\omega^{8}\right]\left(\tilde{\eta}^{2}+\omega^{2}\right)^{-3}\left(\tilde{\nu}^{2}+\omega^{2}\right)^{-2},
\end{align}
%%%%%%%%%%%%%%%%%%%
\begin{align}
\left(\gamma^{(W)}\right)_{b}&=\frac{\chi_{\bar{b}}}{120}\left[4\tilde{\eta}^{2}\left(-3\omega^{4}\tilde{\nu}^{2}+2\omega^{2}\tilde{\nu}^{4}+3\omega^{6}\right)\right.\nonumber\\
-&8\omega^{2}\tilde{\eta}\tilde{\nu}\left(\tilde{\nu}^{2}+\omega^{2}\right)^{2}+\tilde{\eta}^{4}\left(-12\omega^{2}\tilde{\nu}^{2}+3\tilde{\nu}^{4}+\omega^{4}\right)\nonumber\\+
&\left.5\omega^{4}\tilde{\nu}^{4}+11\omega^{8}\right]\left(\tilde{\eta}^{2}+\omega^{2}\right)^{-2}\left(\tilde{\nu}^{2}+\omega^{2}\right)^{-3}\rho^{-1},
\end{align}
%%%%%%%%%%%%%%%%%%%
\begin{align}
\left(\gamma^{(D)}\right)_{u} & =-\frac{\chi_{\rho\bar{u}}}{120}\left[9\tilde{\eta}^{4}\left(\omega^{4}-\tilde{\nu}^{4}\right)+8\tilde{\eta}^{5}\left(5\omega^{2}\tilde{\nu}+6\tilde{\nu}^{3}\right)\right.\nonumber\\
+
 & 8\tilde{\eta}\left(3\omega^{6}\tilde{\nu}+4\omega^{4}\tilde{\nu}^{3}\right)+16\tilde{\eta}^{3}\left(4\omega^{4}\tilde{\nu}+5\omega^{2}\tilde{\nu}^{3}\right)\nonumber\\
 +
 &4\tilde{\eta}^{2}\left(13\omega^{4}\tilde{\nu}^{2}+3\omega^{2}\tilde{\nu}^{4}+10\omega^{6}\right) \nonumber \\
 +& \left.5\omega^{4}\left(4\omega^{2}\tilde{\nu}^{2}+\tilde{\nu}^{4}+3\omega^{4}\right)\right]\nonumber\\ 
 \times
 & \,\left(\tilde{\eta}^{2}+\omega^{2}\right)^{-3}\left(\tilde{\nu}^{2}+\omega^{2}\right)^{-2},
\end{align}
%%%%%%%%%%%%%%%%%%%
\begin{align}
\left(\gamma^{(D)}\right)_{b} & =\frac{\chi_{\bar{b}}}{120}\left[-16\omega^{2}\tilde{\eta}\tilde{\nu}\left(3\omega^{2}\tilde{\nu}^{2}+\tilde{\nu}^{4}+2\omega^{4}\right) \right. \nonumber \\ &\quad+\tilde{\eta}^{4}\left(12\omega^{2}\tilde{\nu}^{2}+19\tilde{\nu}^{4}-23\omega^{4}\right)\nonumber\\ 
-
  8\tilde{\eta}^{3}&\left(3\omega^{4}\tilde{\nu}+4\omega^{2}\tilde{\nu}^{3}+\tilde{\nu}^{5}\right)+\tilde{\eta}^{2}\left(52\omega^{4}\tilde{\nu}^{2}+56\omega^{2}\tilde{\nu}^{4}-36\omega^{6}\right)\nonumber\\ 
 +
 & \left.40\omega^{6}\tilde{\nu}^{2}+37\omega^{4}\tilde{\nu}^{4}-13\omega^{8}\ \right]\left(\tilde{\eta}^{2}+\omega^{2}\right)^{-2}\left(\tilde{\nu}^{2}+\omega^{2}\right)^{-3}\rho^{-1},
\end{align}

\subsection{$\beta$ coefficients}

\begin{equation}
\left(\beta^{(0)}\right)_{u}=\frac{\tilde{\eta}}{3\left(\tilde{\eta}^{2}+\omega^{2}\right)},
\end{equation}
%%%%%%%%%%%%%%%%%%%
\begin{equation}
\left(\beta^{(0)}\right)_{b}=0,
\end{equation}
%%%%%%%%%%%%%%%%%%%
\begin{equation}
\left(\delta^{(\Omega)}\right)_{u}=-\frac{\omega^{2}}{3\left(\tilde{\eta}^{2}+\omega^{2}\right)\left(\tilde{\nu}^{2}+\omega^{2}\right)},
\end{equation}
%%%%%%%%%%%%%%%%%%%
\begin{equation}
\left(\delta^{(\Omega)}\right)_{b}=\frac{\tilde{\nu}^{2}-\omega^{2}}{6\rho\left(\tilde{\nu}^{2}+\omega^{2}\right)^{2}},
\end{equation}
%%%%%%%%%%%%%%%%%%%
\begin{equation}
\left(\delta^{(W)}\right)_{u}=\frac{\tilde{\eta}^{2}-\omega^{2}}{12\left(\tilde{\eta}^{2}+\omega^{2}\right)^{2}},
\end{equation}
%%%%%%%%%%%%%%%%%%%
\begin{equation}
\left(\delta^{(W)}\right)_{b}=\frac{\tilde{\nu}^{2}-\omega^{2}}{12\rho\left(\tilde{\nu}^{2}+\omega^{2}\right)^{2}},
\end{equation}
%%%%%%%%%%%%%%%%%%%
\begin{equation}
\left(\kappa^{(\Omega)}\right)_{u}=\frac{2\omega^{2}\left(11\tilde{\eta}^{2}-5\omega^{2}\right)}{15\left(\tilde{\eta}^{2}+\omega^{2}\right)^{2}\left(\tilde{\nu}^{2}+\omega^{2}\right)},
\end{equation}
%%%%%%%%%%%%%%%%%%%
\begin{equation}
\left(\kappa^{(\Omega)}\right)_{b}=\frac{9\tilde{\nu}^{4}-48\omega^{2}\tilde{\nu}^{2}+7\omega^{4}}{15\rho\left(\tilde{\nu}^{2}+\omega^{2}\right)^{3}},
\end{equation}
%%%%%%%%%%%%%%%%%%%
\begin{align}
\left(\kappa^{(W)}\right)_{u}&= \nonumber \\ &\frac{\tilde{\eta}^{4}\left(23\omega^{2}-\tilde{\nu}^{2}\right)+12\tilde{\eta}^{2}\left(\omega^{4}-\omega^{2}\tilde{\nu}^{2}\right)+5\omega^{4}\left(\tilde{\nu}^{2}+\omega^{2}\right)}{30\left(\tilde{\eta}^{2}+\omega^{2}\right)^{3}\left(\tilde{\nu}^{2}+\omega^{2}\right)},
\end{align}
%%%%%%%%%%%%%%%%%%%
\begin{align}
&\left(\kappa^{(W)}\right)_{b}=\nonumber \\ & \quad\frac{3\tilde{\eta}^{2}\left(-12\omega^{2}\tilde{\nu}^{2}+\tilde{\nu}^{4}+3\omega^{4}\right)-20\omega^{4}\tilde{\nu}^{2}+15\omega^{2}\tilde{\nu}^{4}+13\omega^{6}}{30\rho\left(\tilde{\eta}^{2}+\omega^{2}\right)\left(\tilde{\nu}^{2}+\omega^{2}\right)^{3}},
\end{align}
%%%%%%%%%%%%%%%%%%%
\begin{align}
\left(\beta^{(D)}\right)_{u} & =\frac{1}{30}\left[2\tilde{\eta}^{5}\tilde{\nu}\left(5\tilde{\nu}^{2}+\omega^{2}\right)+16\omega^{2}\tilde{\eta}^{3}\tilde{\nu}^{3}+5\omega^{4}\left(\tilde{\nu}^{2}+\omega^{2}\right)^{2}\right.\nonumber \\
 +& \tilde{\eta}\left(6\omega^{4}\tilde{\nu}^{3}-2\omega^{6}\tilde{\nu}\right)-\tilde{\eta}^{4}\left(10\omega^{2}\tilde{\nu}^{2}+3\tilde{\nu}^{4}+7\omega^{4}\right)\nonumber \\-
 & \left.2\tilde{\eta}^{2}\left(8\omega^{4}\tilde{\nu}^{2}+3\omega^{2}\tilde{\nu}^{4}+5\omega^{6}\right)\right]\left(\tilde{\eta}^{2}+\omega^{2}\right)^{-3}\left(\tilde{\nu}^{2}+\omega^{2}\right)^{-2},
\end{align}
%%%%%%%%%%%%%%%%%%%
\begin{align}
\left(\beta^{(D)}\right)_{b} & =\frac{1}{10}\left[4\tilde{\eta}^{3}\tilde{\nu}^{3}\left(\tilde{\nu}^{2}+\omega^{2}\right)+4\tilde{\eta}^{2}\left(\omega^{6}-3\omega^{4}\tilde{\nu}^{2}\right)\right.\nonumber \\-
 & 4\omega^{2}\tilde{\eta}\tilde{\nu}\left(3\omega^{2}\tilde{\nu}^{2}+\tilde{\nu}^{4}+2\omega^{4}\right)-6\omega^{6}\tilde{\nu}^{2}-\omega^{4}\tilde{\nu}^{4}+3\omega^{8}\nonumber \\+
 & \left.\tilde{\eta}^{4}\left(-6\omega^{2}\tilde{\nu}^{2}+\tilde{\nu}^{4}+\omega^{4}\right)\right]\left(\tilde{\eta}^{2}+\omega^{2}\right)^{-2}\left(\tilde{\nu}^{2}+\omega^{2}\right)^{-3}\rho^{-1},
\end{align}
%%%%%%%%%%%%%%%%%%%
\begin{align}
\left(\kappa^{(D)}\right)_{u} & =\frac{1}{30}\left[2\tilde{\eta}^{5}\tilde{\nu}\left(5\tilde{\nu}^{2}+\omega^{2}\right)+16\omega^{2}\tilde{\eta}^{3}\tilde{\nu}^{3}+\tilde{\eta}\left(6\omega^{4}\tilde{\nu}^{3}-2\omega^{6}\tilde{\nu}\right)\right.\nonumber \\ +
 & \tilde{\eta}^{4}\left(10\omega^{2}\tilde{\nu}^{2}+3\tilde{\nu}^{4}+7\omega^{4}\right)+2\tilde{\eta}^{2}\left(8\omega^{4}\tilde{\nu}^{2}+3\omega^{2}\tilde{\nu}^{4}+5\omega^{6}\right)\nonumber \\ -
 & \left.5\omega^{4}\left(\tilde{\nu}^{2}+\omega^{2}\right)^{2}\right]\left(\tilde{\eta}^{2}+\omega^{2}\right)^{-3}\left(\tilde{\nu}^{2}+\omega^{2}\right)^{-2},
\end{align}
%%%%%%%%%%%%%%%%%%%
\begin{align}
\left(\kappa^{(D)}\right)_{b} & =\frac{1}{30}\left[-4\tilde{\eta}^{3}\tilde{\nu}^{3}\left(\tilde{\nu}^{2}+\omega^{2}\right)+4\omega^{2}\tilde{\eta}\tilde{\nu}\left(3\omega^{2}\tilde{\nu}^{2}+\tilde{\nu}^{4}+2\omega^{4}\right)\right.\nonumber \\ +
  \tilde{\eta}^{4}&\left(-6\omega^{2}\tilde{\nu}^{2}-3\tilde{\nu}^{4}+5\omega^{4}\right)+4\tilde{\eta}^{2}\left(-7\omega^{4}\tilde{\nu}^{2}-4\omega^{2}\tilde{\nu}^{4}+\omega^{6}\right)\nonumber \\ -
 \omega^{4}& \left.\left(22\omega^{2}\tilde{\nu}^{2}+13\tilde{\nu}^{4}+\omega^{4}\right)\right]\left(\tilde{\eta}^{2}+\omega^{2}\right)^{-2}\left(\tilde{\nu}^{2}+\omega^{2}\right)^{-3}\rho^{-1},
\end{align}

\subsection{Helical $\alpha$ coefficients}

\begin{equation}
\left(\tilde{\alpha}_{H}^{(0)}\right)_{u}=\frac{2\tilde{\eta}}{3\left(\tilde{\eta}^{2}+\omega^{2}\right)},
\end{equation}
%%%%%%%%%%%%%%%%%%%
\begin{equation}
\left(\alpha_{H}^{(0)}\right)_{b}=-\frac{2\tilde{\nu}}{3\rho\left(\tilde{\nu}^{2}+\omega^{2}\right)},
\end{equation}
%%%%%%%%%%%%%%%%%%%
\begin{equation}
\left(\gamma^{(\Omega)}\right)_{u}=0,
\end{equation}
%%%%%%%%%%%%%%%%%%%
\begin{equation}
\left(\gamma^{(\Omega)}\right)_{b}=0,
\end{equation}
%%%%%%%%%%%%%%%%%%%
\begin{equation}
\left(\gamma_{H}^{(W)}\right)_{u}=\frac{\tilde{\eta}^{2}\left(\tilde{\nu}^{2}+3\omega^{2}\right)-\omega^{2}\tilde{\nu}^{2}+\omega^{4}}{6\left(\tilde{\eta}^{2}+\omega^{2}\right)^{2}\left(\tilde{\nu}^{2}+\omega^{2}\right)},
\end{equation}
%%%%%%%%%%%%%%%%%%%
\begin{equation}
\left(\gamma_{H}^{(W)}\right)_{b}=\frac{\tilde{\eta}^{2}\left(\omega^{2}-\tilde{\nu}^{2}\right)-\omega^{2}\left(3\tilde{\nu}^{2}+\omega^{2}\right)}{6\rho\left(\tilde{\eta}^{2}+\omega^{2}\right)\left(\tilde{\nu}^{2}+\omega^{2}\right)^{2}},
\end{equation}
%%%%%%%%%%%%%%%%%%%
\begin{align}
\left(\alpha_{H}^{(D)}\right)_{u} & =-\frac{1}{15}\left[3\tilde{\eta}^{4}\left(\omega^{4}-\tilde{\nu}^{4}\right)+4\tilde{\eta}^{5}\left(5\omega^{2}\tilde{\nu}-3\tilde{\nu}^{3}\right) \right.\nonumber \\
  -&\,8\omega^{2}\tilde{\eta}^{3}\tilde{\nu}\left(\tilde{\nu}^{2}-7\omega^{2}\right)+4\omega^{4}\tilde{\eta}\tilde{\nu}\left(\tilde{\nu}^{2}+9\omega^{2}\right) \nonumber \\ 
  +&\,4\tilde{\eta}^{2}\left(11\omega^{4}\tilde{\nu}^{2}+6\omega^{2}\tilde{\nu}^{4}+5\omega^{6}\right)\nonumber \\
  - & \left.5\omega^{4}\left(4\omega^{2}\tilde{\nu}^{2}+\tilde{\nu}^{4}+3\omega^{4}\right)\right]\left(\tilde{\eta}^{2}+\omega^{2}\right)^{-3}\left(\tilde{\nu}^{2}+\omega^{2}\right)^{-2},
\end{align}
%%%%%%%%%%%%%%%%%%%
\begin{align}
\left(\alpha_{H}^{(D)}\right)_{b} & =-\frac{1}{15\rho}\left[\tilde{\eta}^{4}\left(-24\omega^{2}\tilde{\nu}^{2}+7\tilde{\nu}^{4}+\omega^{4}\right) \right. \nonumber \\
-4\tilde{\eta}^{3}&\left(3\omega^{4}\tilde{\nu}+2\omega^{2}\tilde{\nu}^{3}-\tilde{\nu}^{5}\right)
 +4\tilde{\eta}^{2}\left(-11\omega^{4}\tilde{\nu}^{2}+2\omega^{2}\tilde{\nu}^{4}+3\omega^{6}\right)\nonumber \\
 -&4\tilde{\eta}\left(11\omega^{6}\tilde{\nu}+18\omega^{4}\tilde{\nu}^{3}+7\omega^{2}\tilde{\nu}^{5}\right)\nonumber \\
 & \left.+\omega^{4}\left(-20\omega^{2}\tilde{\nu}^{2}+\tilde{\nu}^{4}+11\omega^{4}\right)\right]\left(\tilde{\eta}^{2}+\omega^{2}\right)^{-2}\left(\tilde{\nu}^{2}+\omega^{2}\right)^{-3}.
\end{align}
All of the listed kinematic transport coefficients agree with those
given in \citetalias{Radler:2006hy}, with one exception. This is the $\left(\beta^{(D)}\right)_{u}$
coefficient, which contains a factor $1/30$, rather than $1/60$.

\section{The sign of $(\eta_{yx})_{b}^{S}$}\label{sec:The-sign-of etab}

In this appendix we give argue that the sign of $(\eta_{yx})_{b}^{S}$
is always negative, given reasonable assumptions about the form of
$W_{b}\left(k,\omega\right)$. We have not been able to find a general
proof that this is the case due to the complexity of the expression
Eq.~\eqref{eq: eta Sb}, but instead analyze the cases $\mathrm{Pm}=1$,
$\mathrm{Pm}\ll1$, and $\mathrm{Pm}\gg1$ separately. In addition,
plotting $(\eta_{yx})_{b}^{S}$ for Gaussian $W_{b}$ {[}Eq.~\eqref{eq:gaussian W}{]}
across a range of $\mathrm{Pm}$ (e.g., Fig.~\ref{fig:Transport-coefficients-S}) leads us to the same
conclusion for this specific $W_{b}$. [Note that $(\eta_{yx})_{b}^{S}$
depends nontrivially only on $\mathrm{Pm}$ and $q$ when written
in the dimensionless variables given in Eq.~\eqref{eq:DIMENSIONLESS variables},
meaning it is straightforward to observe positivity by plotting $(\eta_{yx})_{b}^{S}$
against $q$ over a range of $\mathrm{Pm}$.]

\subsection{$\mathrm{Pm}=1$}

Inserting $\nu=\eta$ into Eq.~\eqref{eq: eta Sb} leads
to 
\begin{equation}
(\eta_{yx})_{b}^{S}=\int d\omega\, dk\, k^{2}W_{b}\left(k,\omega\right)\frac{8\pi\left(\omega^{2}-\tilde{\eta}^{2}\right)\left(3\tilde{\eta}^{2}+\omega^{2}\right)}{15\left(\tilde{\eta}^{2}+\omega^{2}\right)^{3}}.
\end{equation}
An integration by parts in $\omega$ yields 
\begin{align}
(\eta_{yx})_{b}^{S} & =\frac{4\pi}{15}\int d\omega \,dk\,\left[\frac{1}{\eta}\tan^{-1}\left(\frac{\omega}{\tilde{\eta}}\right)\frac{dW_{b}}{d\omega}\right.\nonumber \\
 &\qquad\qquad\qquad+ \left.\frac{5\tilde{\eta}^{2}+3\omega^{2}}{\left(\tilde{\eta}^{2}+\omega^{2}\right)^{2}}\omega\frac{dW_{b}}{d\omega}\right].\label{eq:eta pos Pm1}
\end{align}
Under the reasonable assumptions that $\omega\, dW/d\omega\leq0$
and $\tan^{-1}\left(\omega\right)dW/d\omega\leq0$, each term in the
integral must be negative. (Note that the $\tan^{-1}\left(\omega\right)dW/d\omega\leq0$ condition,
although it may appear less familiar, is just as restrictive as $\omega\, dW/d\omega\leq0$,
given the odd nature of the $\tan^{-1}$ function.)

\subsection{$\mathrm{Pm}\ll1$}

Inserting $\eta=\nu/\mathrm{Pm}$ into Eq.~\eqref{eq: eta Sb},
we carry out a series expansion about $\mathrm{Pm}^{-1}=\infty$ of
the resulting expression. The reason for this expansion (rather than
the more obvious expansion about $\mathrm{Pm}=0$) is that we wish
to explore that low $\mathrm{Pm}$ limit with large $\eta$
rather than that with $\nu\rightarrow0$, since  SOCA looses
applicability as $\nu,$$\eta\rightarrow0$. The series
expansion to first order in $1/\mathrm{Pm}^{-1}$ is 
\begin{align}
(\eta_{yx})_{b}^{S} & \approx-\frac{8\pi}{15}\int d\omega dk\, W_{b}k^{2}\left(\frac{3\omega^{2}\tilde{\nu}^{2}\tilde{\nu}^{4}-2\omega^{4}}{\left(\tilde{\nu}^{2}+\omega^{2}\right)^{3}}\right.\\
 & \qquad\qquad+\left.\frac{4\tilde{\nu}^{2}}{15\left(\tilde{\nu}^{2}+\omega^{2}\right)^{2}}\frac{1}{\mathrm{Pm}^{-1}}+\dots\right).
\end{align}
The first term is independent of $\mathrm{Pm}$, persisting as $\eta\rightarrow0$,
and the existence of this is not surprising given the fact that the
dynamo can arise from the $\bm{B}\cdot\nabla\bm{b}+\bm{b}\cdot\nabla\bm{B}$
term in the induction equation. This term can be shown to be negative
using the same integration by parts method used to obtain Eq.~\eqref{eq:eta pos Pm1},
with the requirement $\omega\, dW/d\omega\leq0$. The $\mathrm{Pm}$
dependent second term is obviously negative due to the positive definiteness
of the integrand.

\subsection{$\mathrm{Pm}\gg1$}

Inserting $\nu=\mathrm{Pm}\,\eta$ into Eq.~\eqref{eq: eta Sb},
and carrying out a series expansion about $\mathrm{Pm}=\infty$ (see
previous section), one obtains 
\begin{equation}
(\eta_{yx})_{b}^{S}\approx\frac{16\pi}{15}\int d\omega dk\, W_{b}k^{2}\left(\frac{1}{\mathrm{Pm}}\frac{\left(\omega^{2}-\tilde{\eta}^{2}\right)}{\left(\tilde{\eta}^{2}+\omega^{2}\right)^{2}}+\dots\right).
\end{equation}
As expected, there is no $\nu=0$ contribution to the transport.
Again using integration by parts, it is easy to prove negativity of the
integral provided $\omega\, dW/d\omega\leq0$.
\end{document}